\documentstyle[eqsecnum,prb,aps]{revtex}
\input epsf
\begin{document}
\draft
\title{Averaged Methods for Vortex-String Evolution}
\author{C. J. A. P. Martins\thanks{Also at C. A. U. P.,
Rua do Campo Alegre 823, 4150 Porto, Portugal.
Electronic address: C.J.A.P.Martins\,@\,damtp.cam.ac.uk}
and
E. P. S. Shellard\thanks{Electronic address:
E.P.S.Shellard\,@\,damtp.cam.ac.uk\vskip0pt Submitted to
Phys.\ Rev.\ {\bf B}.}}
\address{Department of Applied Mathematics and
Theoretical Physics\\
University of Cambridge\\
Silver Street, Cambridge CB3 9EW, U.K.}
\maketitle

\begin{abstract}
We discuss friction-dominated vortex-string evolution
using a new analytic model recently developed by the
authors. By treating the average string velocity, as
well as the characteristic lengthscale, as dynamical
variables, we can provide a quantitative picture of the
complete evolution of a vortex-string network.
Previously known scaling laws are confirmed, and new
quantitative predictions regarding loop production and
evolution are made.
\end{abstract}

\section{Introduction}

The concept of symmetry breaking plays a crucial role in
modern physics, and
one of its most interesting consequences
is the formation of topological defects. These defects
have been observed and studied in a wide variety of
condensed matter contexts, including metal
crystallization\cite{metc},
liquid crystals\cite{cdty,lctw}, superfluid
helium\cite{msgv,zurek} and superconductors\cite{abrk}.
In models where they are allowed, defects will form
whenever the rate of the phase transition is fast
relative to the
scale of the system size (in other words, a `quench').

On the other hand, they are also believed to have formed
in the early universe, and they can play an extremely
important part in its evolution\cite{vs}. In this
context, the conditions for their formation were first
established by Kibble\cite{kib0}---except for some
subtleties in the case of the breaking of a gauge
symmetry\cite{gauge},
they are entirely analogous.

The scaling evolution of vortex-string networks has been
extensively studied analytically in both condensed
matter and cosmological settings\cite{k85,ack,ab},
but using rather different methods. This difference is
perhaps understandable given the extremity of these two
physical regimes, but it may not be necessary. Condensed
matter descriptions tend to focus on a coarse-grained
order parameter $\phi$, providing a low-level picture of
defect motion by estimating energy dissipation rates. On
the other hand, high energy physicists take an
`idealized' one-dimensional view of string dynamics
by integrating out the radial degrees of freedom (in the
Higgs $\phi$ and other fields) to obtain a low-energy
effective action---the Nambu action. The resulting
relativistic equations of motion can then be averaged to
describe the large-scale evolution of the string
network. Naturally one should also account for energy
loss mechanisms, such as loop production---something
that is not done in condensed-matter contexts.

The analytical study of cosmological string networks was
started one decade ago by Kibble's `one-scale'
model\cite{k85} (later modified by Bennett\cite{b1}
and Albrecht \& Turok\cite{albrecht}). There it was
assumed that the evolution of the long-string network
could be described by a single lengthscale, which is
usually called the `correlation length'. One then
supposes that a scaling solution exists at late times
and ends up showing that such a solution will in fact
exist and be stable subject to conditions on the
loop production mechanisms. One of the caveats of this
model is that it is only valid in the latter
evolutionary stages when strings are moving
relativistically. However, at early times the string
dynamics is friction-dominated, due to
particle--string scattering. Of course, this is also
the relevant case in condensed matter physics.

A model of string evolution including the effects of
frictional forces has been recently proposed by the
authors\cite{ms,ms1}. It is a simple generalization of
the `one-scale' model in which the average string rms
velocity becomes a dynamical variable. This simple model
provides the first quantitative description of the
complete evolution of a string network in the early
universe\cite{ms,ms1}. In this paper, we will use it to
study string evolution at constant temperature, which is
relevant in condensed matter contexts. The main
advantage of this approach is that one can easily
account for the production and evolution of string
loops. Although it should be seen as the basis for
further work, the model is already predictive enough to
be testable in laboratory experiments.

The structure of this paper is as follows. In the next
section, after a review of string dynamics, the
evolution equations for the `characteristic lengthscale'
and the average velocity of the long string network and
each individual loop are derived and justified. The
cases of strings arising from the breaking of gauge and
global symmetries are both considered. The validity of
these `averaged' evolution equations is then tested
against a simple---but physically relevant---loop
solution in section 3.
In section 4 we describe string evolution at constant
temperature. The case where the dynamics is
friction-dominated is directly applicable to a whole
range of situations in condensed-matter systems. It will
be shown that this model can easily reproduce some
well-known results, notably the $L\propto t^{1/2}$ law.
Furthermore, new quantitative predictions regarding loop
evolution are made. We also discuss the opposite limit,
where strings are initially `free' and find that in that
case a network quickly evolves towards the previous
regime. We note that the evolution equations for
the string network and the string loops are
relatively straightforward, but the intervening
derivation is not. First-time readers may wish to
skip straight to the results in section IV.

\section{The averaged evolution equations}
\subsection{Basics of string dynamics}
The usual condensed matter approach to vortex
dynamics is based on a `coarse-grained'
complex scalar field $\phi$. In quantum field
theory, for example, one can consider the
abelian-Higgs model, which is a relativistic
generalization of the Ginzburg-Landau theory of
superconductivity. It is also of interest to
consider the global version of this, that
is the Goldstone model.

In high-energy physics it also proves to be 
convenient to adopt a one-dimensional view of
string dynamics (see \cite{vs} for a detailed
discussion of the issues below). In this description
a string sweeps out a two-dimensional surface (the
worldsheet) which can be described by two
`worldsheet coordinates'---one is time-like and can
in fact be identified with the background time
(which we will denote by $t$), while the other is
space-like and simply labels points along the string
(we will call it $\sigma$).

This one-dimensional description is essentially
acheived by integrating over the radial modes of the 
vortex solution on the assumption that the
scale of perturbations along the string is much
larger than its width---thereby
obtaining a low-energy effective action.
For the case of a gauge (global) string,
one thereby obtains the Nambu (Kalb-Ramond) action
from the abelian-Higgs (Goldstone) model.
Note that in the global case there are long range
forces between the strings, in general there
can be both external and self-field contributions
to this. However, it can be shown (again, see
\cite{vs} for a detailed discussion) that the later
can be `renormalized out', yielding the
well-known logarithmically divergent string energy per
unit length $\mu$ (which is a constant for gauge
strings). By varying these actions it is then
straightforward to obtain the string equations
of motion.

There is, however, a crucial ingredient for
string evolution missing. Since strings move through
a background fluid, their motion is retarded by particle
scattering. Vilenkin has shown\cite{v1} that this effect
can be described by a frictional force per unit length
that can be written
\begin{equation}
{\bf F}_{\rm f}=-\frac{\mu}{\Gamma}\gamma{\bf v}
\, , \label{ff}
\end{equation}
where ${\bf v}$ is the string velocity, $\gamma$ is the
Lorentz factor and $\Gamma$ is a constant
damping coefficient, that can be written as
the square of a characteristic propagation speed (which
need not necessarily be the speed of light)
times a `friction timescale' $\tau_{\rm f}$, whose
explicit value depends on the type of
symmetry involved. For a gauge string, the main
contribution comes from Aharonov-Bohm
scattering\cite{rsa}, while in the global case it comes
from Everett scattering\cite{epa}. For example, if the
background fluid is a perfect gas, we have
\begin{equation}
\tau_{\rm f}=\left\{ \begin{array}{ll}
\frac{2\pi\hbar}{\beta}
\frac{(k_B T_c)^2}{(k_B T)^3} & \mbox{Gauge} \\
\frac{2\pi\hbar}{\beta}\frac{(k_B T_c)^2}{(k_B T)^3}
\ln\left(\frac{R}{\delta}\right)
\ln^2(T\delta) & \mbox{Global}
 \end{array} \right.
\label{frilen}
\end{equation}
where $T$ is the background temperature and $\beta$ is
a numerical factor related to the number of particle
species interacting with the string (strictly speaking,
its value is slightly different in the two cases, but a
common symbol will be used for simplicity). It should
also be noted that the Everett scattering formula is
only valid when the particle wavelength is much larger
than the string thickness $\delta$. Alternatively,
$\Gamma$ can be written as the characteristic propagation
speed times a `friction lengthscale' $\ell_{\rm f}$.

For string motion in a flat background,
the string equations of motion with the frictional force
(\ref{ff}) can then be written as\cite{ms1}
\begin{equation}
\frac{1}{c^2}{\ddot{\bf x}}+
\left(1-\frac{{\dot{\bf x}}^2}{c^2}\right)
\frac{\dot{\bf x}}{\Gamma}=\frac{1}{\epsilon}
\left(\frac{{\bf x}'}{\epsilon}\right)'
\, , \label{spc}
\end{equation}
\begin{equation}
{\dot\epsilon}+\frac{{\dot{\bf x}}^2}{\Gamma}
\epsilon=0
\, , \label{tc}
\end{equation}
where the dimensionless parameter $\epsilon$ (which can be
interpreted as a `coordinate energy per unit length')
is defined by
\begin{equation}
{\epsilon^2=}\frac{{{\bf x}'}^2}{{1-{\dot{\bf x}}^2/c^2}}
\, , \label{epsil}
\end{equation}
and dots and primes respectively denote time and
space derivatives. This form of
the evolution equations proves to be particularly
useful because dissipation is naturally incorporated in
the decay of the coordinate energy density $\epsilon$,
while preserving the gauge conditions. Note that
while this is a truly relativistic formalism, it is
straightforward to obtain the non-relativistic limit
that will be adequate to condensed matter contexts
where the dynamics is friction-dominated.
In this case, Eqns. (\ref{spc}-\ref{epsil})
reduce to
\begin{equation}
\frac{{\dot{\bf x}}}{\Gamma}=-\frac{1}{{{\bf x}'}^4}
\left[{\bf x}'\wedge\left({\bf x}'
\wedge{\bf x}''\right)\right]
\, , \label{nrll}
\end{equation}
and one can recognize the right-hand side as the friction
force term (which is dominant in this limit),
e.g. on a superfluid vortex\cite{schwarz}

Incidentally, it has been shown\cite{dsmag} that a
global string will behave as a superfluid vortex if it
is introduced in a homogeneous background (with a
density $\rho_h$, say)---physically, this corresponds
to giving it angular
momentum. The interaction between this background and
the string originates an additional force, known as the
(relativistic) Magnus force, and (\ref{spc}) becomes
\begin{equation}
\frac{1}{c^2}{\ddot{\bf x}}+\left(1-\frac{{\dot{\bf
x}}^2}{c^2}\right)\frac{\dot{\bf x}}{\Gamma}=\frac{1}{\epsilon}
\left(\frac{{\bf x}'}{\epsilon}\right)'+
\Gamma'\dot{\bf x}\wedge\frac{{\bf x}'}{\epsilon}
\, , \label{magnus}
\end{equation}
where $\Gamma'\propto \rho_h^{1/2}$;
note that (\ref{tc}) remains unchanged.

\subsection{Lengthscale evolution}
We can now proceed to average the string equations of
motion to describe the large-scale evolution of the
string network. We therefore define the total string
energy and the average rms string velocity to be
\begin{equation}
E=\mu\int\epsilon d\sigma
\, , \label{et}
\end{equation}
\begin{equation}
v^2\equiv \langle{\dot{\bf x}}^2\rangle=
\frac{{\int{\dot{\bf x}}^2\epsilon d\sigma}}
{{\int\epsilon d\sigma}}
\, . \label{vv}
\end{equation}
Differentiating (\ref{et}) and using (\ref{tc}) and
(\ref{vv}), we see that the total string energy density
will obey the following evolution equation:
\begin{equation}
\frac{d\rho}{dt}+\frac{v^2}{\Gamma}\rho=0
\, . \label{dens}
\end{equation}

This includes both long strings and loops which have a
low probability of interacting with other strings before
decaying. We shall study the evolution of the
long-string network on the assumption that it can be
characterized by a single lengthscale $L$; this can be
interpreted as the inter-string distance or the
`correlation length'. Strings larger than $L$ will be
called long or `infinite'; otherwise they will be called
loops. For Brownian long strings, we can define the
`correlation length' $L$ in terms of the network
density\footnote{Throughout this paper the subscript
`$\infty$' refers to properties of the long (`infinite')
string network.} $\rho_\infty$ as
\begin{equation}
\rho_{\infty}\equiv\frac{\mu}{L^2}
\, . \label{correlation}
\end{equation}

Following Kibble\cite{k85}, the rate of loop production
from long-string collisions can be estimated as follows.
Conceptually, we divide the network into a collection of
segments of length $L$, each in a volume $L^3$. Consider
another segment of length $l$ moving with a velocity
$v_\infty$; the probability of it encountering one of
the other segments within a time $\delta t$ is
approximately $lv_\infty\delta t/ L^2$. Consistently
with our `one-scale' assumption, we then assume that the
probability of such an intersection creating a loop of
length in the range $l$ to $l+dl$ will be given by a
scale-invariant function $w\left({l/L}\right)$. The rate
of energy loss into loops is then given by
\begin{equation}
\left(\frac{d\rho_{\infty}}{dt}\right)_{\rm to\ loops}=
\rho_{\infty}\frac{v_\infty}{L}
\int w\left(\frac{\ell}{L}\right)\frac{\ell}{L}
\frac{d\ell}{L}\equiv {\tilde
c}v_\infty\frac{\rho_{\infty}}{L}
\, . \label{rtl}
\end{equation}
where the loop `chopping' efficiency ${\tilde c}$ is
assumed to be constant. By subtracting the loop energy
losses (\ref{rtl}) from (\ref{dens}) and then using
(\ref{correlation}), we obtain the overall evolution
equation for the characteristic lengthscale $L$,
\begin{equation}
2 \frac{dL}{dt}=\frac{v_\infty^2}{\Gamma}L+
{\tilde c}v_\infty
\, . \label{evl}
\end{equation}

\subsection{Loop evolution}
The main advantage of the present approach is that we
can also study the evolution of the loop density and
distribution, which have so far been neglected in the
condensed matter literature. The traditional approach
in cosmology is to define $n_\ell(\ell,t)d\ell$ to be
the number density of loops with length in the range
$(\ell,\ell+d\ell)$ at time $t$; the corresponding loop
energy density distribution is
\begin{equation}
\rho_{\ell}(\ell,t)d\ell = \mu\ell n_\ell(\ell,t)d\ell
\, . \label{ledd}
\end{equation}
Note that the total loop energy density is
\begin{equation}
\rho_{o}\equiv\int\rho_{\ell}(\ell,t)d\ell
\, , \label{leddo}
\end{equation}
and $\rho=\rho_{\infty}+\rho_{o}$\footnote{Throughout
this paper the subscript `$o$' refers to properties of
the entire loop population, while `$\ell$' refers to the
loops with length in the range $(\ell,\ell +d\ell)$.}
{}From our assumptions on the loop production rate
(\ref{rtl}) it is then easy to see that
\begin{equation}
\frac{d\rho_{\ell}}{dt}+
\frac{v_{\ell}^2}{\Gamma}\rho_{\ell}=\mu
\frac{v_\infty\ell}{L^5}w \left(\frac{\ell}{L}\right)
\, . \label{evledd}
\end{equation}
However, note that this equation is `static', in the
sense that it does not include loop decay mechanisms.

Instead, we start by using our analytic model to describe
the evolution of each individual loop. Knowing the energy
density transferred from long strings into loops and
estimating their sizes at formation (see below), one can
determine the energy density in loops and other relevant
quantities at all times. This formalism does not allow
for possible loop reconnections or self-intersections
but these should not be important in a
friction-dominated regime.

The physical size of a loop is simply given by
\begin{equation}
\ell=\int_{loop}\epsilon d\sigma
\, , \label{physl}
\end{equation}
and its time evolution is easily found to be
\begin{equation}
\frac{d\ell}{dt}=-\ell
\frac{v^2_\ell}{\Gamma} \, . \label{evlps}
\end{equation}

Now, we will assume that loop production is
`monochromatic', ie that loops formed at a time
$t_p$ have an initial length
\begin{equation}
\ell(t_p)=\alpha(t_p)\ L(t_p)
\, , \label{inls}
\end{equation}
where $\alpha$ is a parameter of order unity.

With this ansatz the scale-invariant loop production
function $w$ (see (\ref{rtl})) becomes
\begin{equation}
w\left(\frac{\ell}{L}\right)=\frac{{\tilde c}}{\alpha}
\delta\left(\frac{\ell}{L}-\alpha\right)
\, , \label{wansatz}
\end{equation}
and the rate of energy loss into loops becomes
\begin{equation}
\left(\frac{d\rho_{\infty}}{dt}\right)_{\rm to\ loops}=
\mu{\tilde c}\frac{v_\infty}{L^3}
\, . \label{rtlalpha}
\end{equation}

Hence the energy density converted into loops from time
$t$ to $t+dt$ is
\begin{equation}
d\rho_o(t)=\mu{\tilde c}\frac{v_\infty}{L^3}dt
\, ; \label{rozdt}
\end{equation}
this corresponds to a fraction
\begin{equation}
\frac{d\rho_o(t)}{\rho_{\infty}(t)}={\tilde c}
\frac{v_\infty}{L}dt
\, , \label{corfr}
\end{equation}
of the energy density in the form of long strings at
time $t$. Then using our ansatz (\ref{wansatz}), the
corresponding number of loops produced in a volume $V$ is
\begin{equation}
dN(t)=\frac{{\tilde c}}{\alpha}\frac{v_\infty}{L^4}Vdt
\, ; \label{numlo}
\end{equation}
hence the ratio of the energy densities in loops and
long strings at time $t$ is (neglecting
the initial loop population at $t=t_c$)
\begin{equation}
\varrho(t)\equiv\frac{\rho_o(t)}{\rho_{\infty}(t)}=L^2(t)
\int_{t_c}^{t}\frac{dN(t')\ell(t,t')}{V}={\tilde c}L^2(t)
\int_{t_c}^{t}\frac{v_\infty (t')}{L^4(t')}
\frac{\ell(t,t')}{\alpha (t')}dt'
\, , \label{ratio}
\end{equation}
where $t_c$ is the moment of the network formation and
$\ell(t,t')$ is the length at time $t$
of loops produced at time $t'$.

We can therefore numerically (and, in some simple limit
cases, analytically) determine the loop density at all
times. This generalized `one-scale' model can therefore
provide a complete description of a string network.

\subsection{Velocity evolution}
We now consider the evolution of the average string
velocity $v$. A non-relativistic equation can be easily
obtained: it is just Newton's law,
\begin{equation}
\frac{\mu}{c^2}\frac{dv}{dt}=\frac{\mu}{R}-\mu
\frac{v}{\Gamma}
\, . \label{nonrelvel}
\end{equation}
This merely states that curvature accelerates the strings
while friction slows them down. On dimensional grounds,
the force per unit length due to curvature should be
$\mu$ over the curvature radius $R$. The form of the
damping force can be found similarly.

A relativistic generalization of the velocity evolution
equation (\ref{nonrelvel}) can be obtained more
rigorously by differentiating (\ref{vv}):
\begin{equation}
\frac{1}{c^2}\frac{dv}{dt}=\left(1-\frac{v^2}{c^2}\right)
\left(\frac{k}{R}-\frac{v}{\Gamma}\right)
\, . \label{evv}
\end{equation}

This is exact up to second-order terms. In the curvature
term, we have introduced $R$ via the definition of the
curvature radius vector,
\begin{equation}
\frac{\hat{\bf u}}{R}=\frac{d^2{\bf x}}{ds^2}
\, , \label{crv}
\end{equation}
where $\hat{\bf u}$ is a unit vector and $s$ is the
physical length along the string (related to the
coordinate length $\sigma$ by $ds=|{\bf x}'|d\sigma =
\left(1-{\dot{\bf x}}^2/c^2\right)^{1/2} \epsilon d\sigma$).
 The dimensionless parameter $k$ is defined by
\begin{equation}
\langle(1-\frac{{\dot{\bf x}}^2}{c^2})({\dot{\bf x}}\cdot
\hat{\bf u})\rangle\equiv kv\left(1-\frac{v^2}{c^2}\right)
\, . \label{dfk}
\end{equation}

Note that in the case of long strings, our `one-scale'
assumption implies that the curvature radius coincides
with the correlation length, $R\equiv L$; on the other
hand, for a loop of size $\ell$ we should have
$\ell\approx 2\pi R$.

The parameter $k$ is `phenomenological', and is related
to the presence of small-scale `wiggles' on strings: on
a perfectly smooth string, $\hat{\bf u}$ and
${\dot{\bf x}}$ will be parallel so $k=1$ (up to a
second-order term as above); however this need not be
so for a wiggly string\cite{ms1}.
On the other hand, one can show
that in flat spacetime (with no friction) $k=0$.

In the case of long strings, we should expect $k=1$
since the correlation length $L$ is much larger than
the `friction length' $\ell_{\rm f}$.
Now, consider a particular string
loop. While it is large compared with the friction
lenghtscale, taking $k\approx1$ should again be a good
assumption. In the opposite limit friction is no longer
effective in damping its motion, and it moves as if it
was in flat spacetime. Thus one requires that
$k\rightarrow0$ as $R\rightarrow0$. In particular,
demanding that their limiting velocity be
$v^2_\ell/c^2=1/2$ leads to the requirement that
$k\propto R$ as $R\rightarrow0$. With these requirements
in mind, and after comparing with the `microscopic'
(ie, unaveraged) evolution of some simple solutions
(to be described in the next section)
one arrives at the following ansatz:
\begin{equation}
k=\left\{ \begin{array}{ll}
1\, , & \mbox{$\frac{R}{\ell_{\rm f}}>\chi$} \\
\frac{1}{\sqrt2}\frac{R}{\ell_{\rm f}}\, , &
\mbox{$\frac{R}{\ell_{\rm f}}<\chi$}
 \end{array} \right.
\label{kans}
\end{equation}
where $\chi$ is a numerical coefficient\cite{gs} of
order one. Recall that the physical loop length is
approximately $\ell=2\pi R$, while for the
long-string network $L=R$.

Equations (\ref{evl}), (\ref{evlps}) and (\ref{evv})
form the basis of our generalized `one-scale' model,
which we will now proceed to apply.

\section{`Averaged' versus `microscopic' evolution}
In order to check the validity of our `averaged'
evolution model, and in particular our ansatz for $k$,
we will test it against a simple loop solution.

Consider a circular loop in a flat background with a
constant friction timescale---that is,
a condensed-matter-like situation. We can describe
the loop trajectory simply by
\begin{equation}
{\bf x}=r(t) (\sin\theta,\cos\theta,0)\, , \,
\theta\in[0,2\pi]
\, . \label{ccl}
\end{equation}
Then equations (\ref{spc},\ref{tc}) reduce to
\begin{equation}
\frac{1}{c^2}{\ddot r}+(1-\frac{{\dot r}^2}{c^2})
\left(\frac{{\dot r}}{\Gamma}+
\frac{1}{r}\right)=0
\, . \label{smallr}
\end{equation}
Note that the physical (`invariant') loop radius is
$R=r/\sqrt{1-{\dot r}^2/c^2}$, obeying
\begin{equation}
{\dot R}=-R\frac{{\dot r}^2}{\Gamma} \, ; \label{bigr}
\end{equation}
also the `microscopic' velocity is $v=-{\dot r}$ and
obeys
\begin{equation}
\frac{1}{c^2}{\dot v}=\left(1-\frac{v^2}{c^2}\right)
\left(\frac{1}{r}-\frac{v}{\Gamma}\right)
\, . \label{unavv}
\end{equation}

On the other hand, our averaged evolution equations
(\ref{evlps},\ref{evv}) take the form
\begin{equation}
\frac{d{\bar R}}{dt}=-{\bar R}
\frac{{\bar v}^2}{\Gamma} \, , \label{avevr}
\end{equation}
\begin{equation}
\frac{1}{c^2}\frac{d{\bar v}}{dt}=\left(1-
\frac{{\bar v}^2}{c^2}\right)\left(
\frac{k({\bar R})}{{\bar R}}-
\frac{{\bar v}}{\Gamma}\right) \, . \label{avevrv}
\end{equation}
Notice the similarity between the two approaches. Loops
with size much larger than a `friction lengthscale' that
can be defined in the obvious way from the friction timescale
will be overdamped, with the velocity being approximately
given by
\begin{equation}
v\sim\frac{\Gamma}{r}
\, . \label{fovd}
\end{equation}
In this case the two sets of evolution equations actually
coincide---hence justifying our $k=1$ ansatz for large R.
As the loop gains velocity $r$ and $R$ become
significantly different and this equivalence ceases to
be valid. When $R$ becomes much
smaller than $\ell_{\rm f}$,
the loop still looses energy due to friction, but this
is no longer effective in damping its motion---the loop
now begins to oscillate relativistically. In particular,
over one `period' $v$ oscillates between $0$ and
$1$. But we know that the averaged velocity should (in
the limit) be ${\bar v}^2/c^2=1/2$; this is the physical
reason why we need a `phenomenological' varying $k$ on
small scales. As we mentioned previously, this
requirement fixes the behaviour of $k$ on small scales
to be as shown in (\ref{kans}). The remaining question
is then how to match the two regimes.

First of all, we need a clear idea of when (and where)
the transition occurs. A good guess would be the moment
of the `first collapse', ie, the moment when we first
have $v=c$. In fact, this turns out to be a well-defined
event. As was first pointed out by Garriga and
Sakellariadou\cite{gs} (and can be easily seen by
analytical or numerical study of the equation of motion
(\ref{smallr})), circular loops with initial radius much
larger than the friction length always reach $v=c$ for
the first time when
\begin{equation}
\left(\frac{R}{\ell_{\rm f}}\right)_{col}=\chi_{c}
\simeq0.5691\, . \label{defchi}
\end{equation}
Note that $r_i\gg\ell_{\rm f}$
is the physically relevant case
for string dynamics in condensed matter contexts (recall
that the dynamics in that case is always
friction-dominated). Also note that because of friction,
all loops will rapidly become (almost) circular.

\begin{figure}
\vskip-0.1in
\vbox{\centerline{
\epsfxsize=.8\hsize\epsfbox{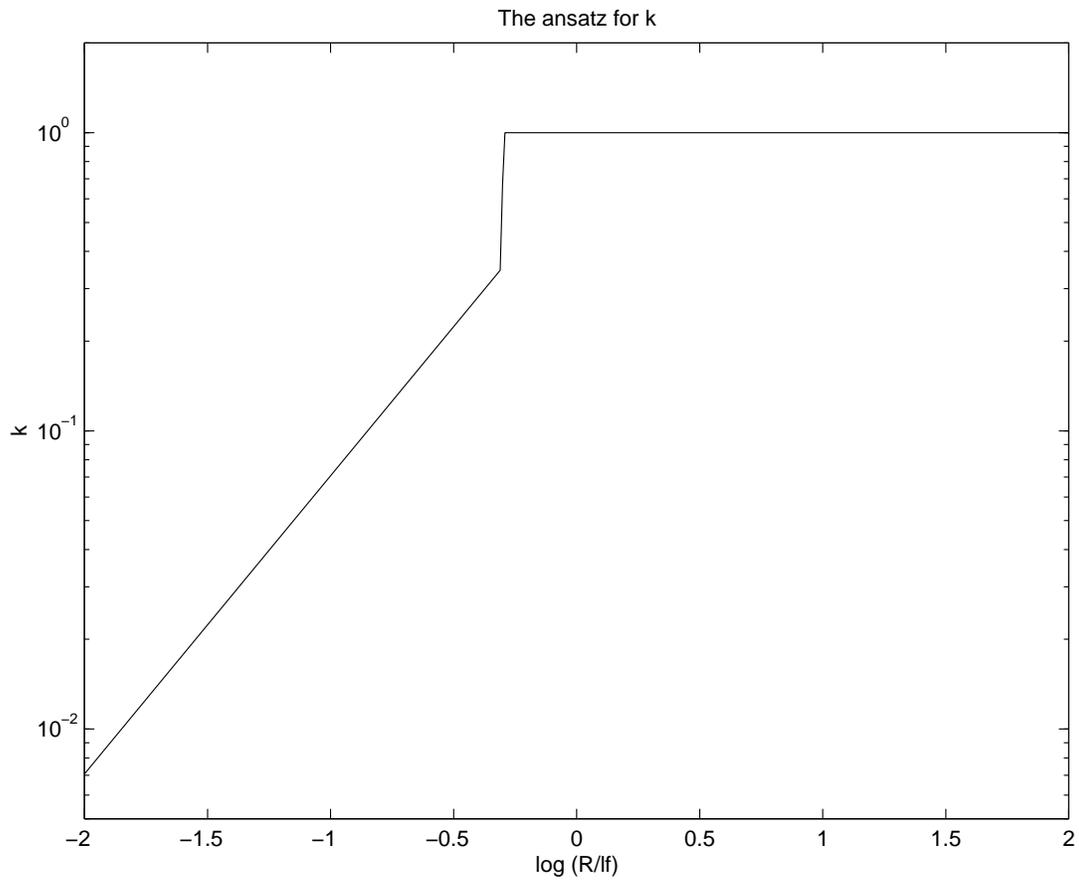}}
\vskip.4in}
\caption{The ansatz for parameter $k$ for string loops
in a condensed-matter context as a function of the
rescaled loop radius. The behaviour at large and small
scales has a physical justification (see text); matching
was done numerically.}
\label{fig31}
\end{figure}

\begin{figure}
\vskip-1in
\vbox{\centerline{%
\hskip-3em\epsfxsize=.5\hsize\epsfbox{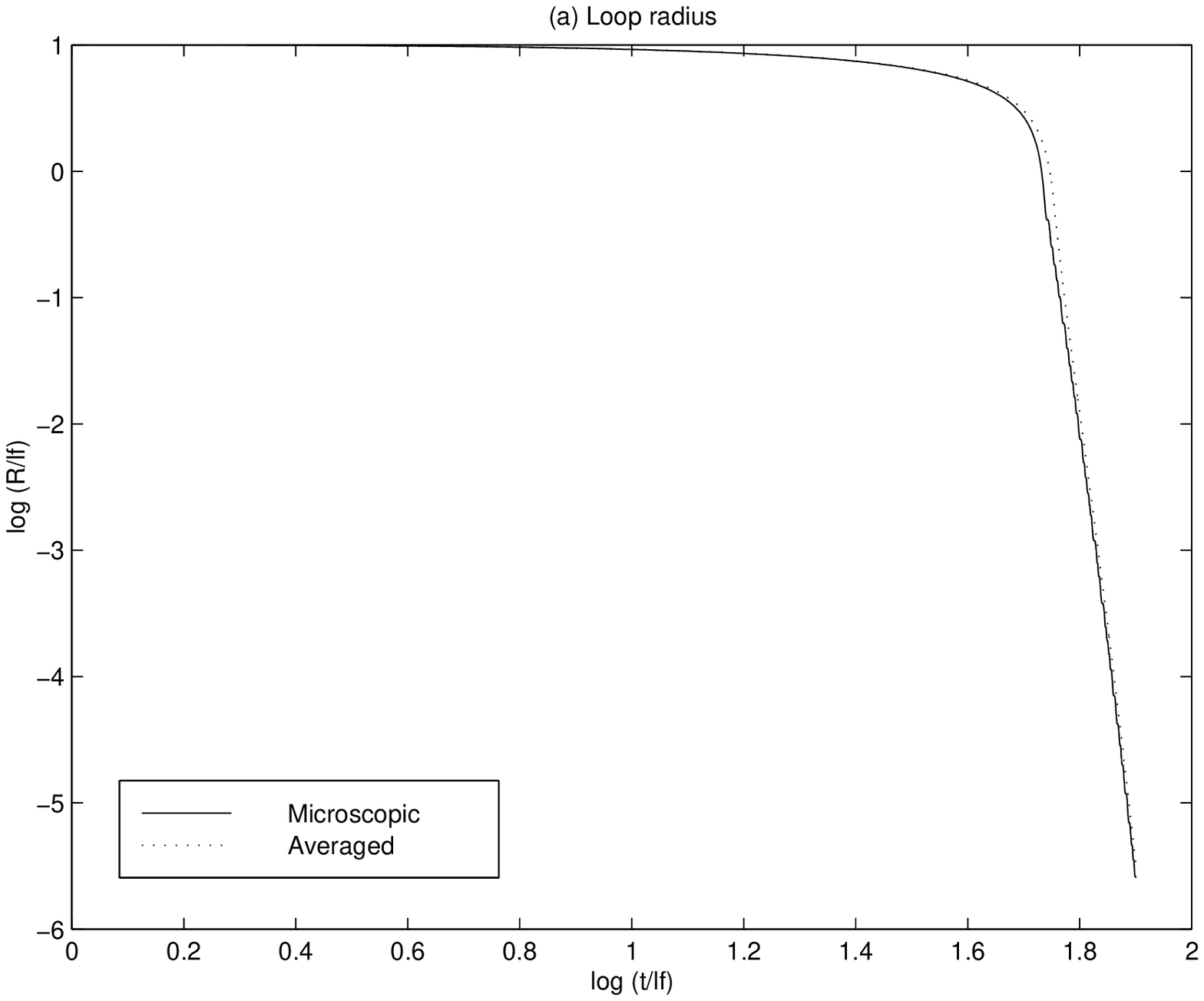}
\hskip3em\epsfxsize=.5\hsize\epsfbox{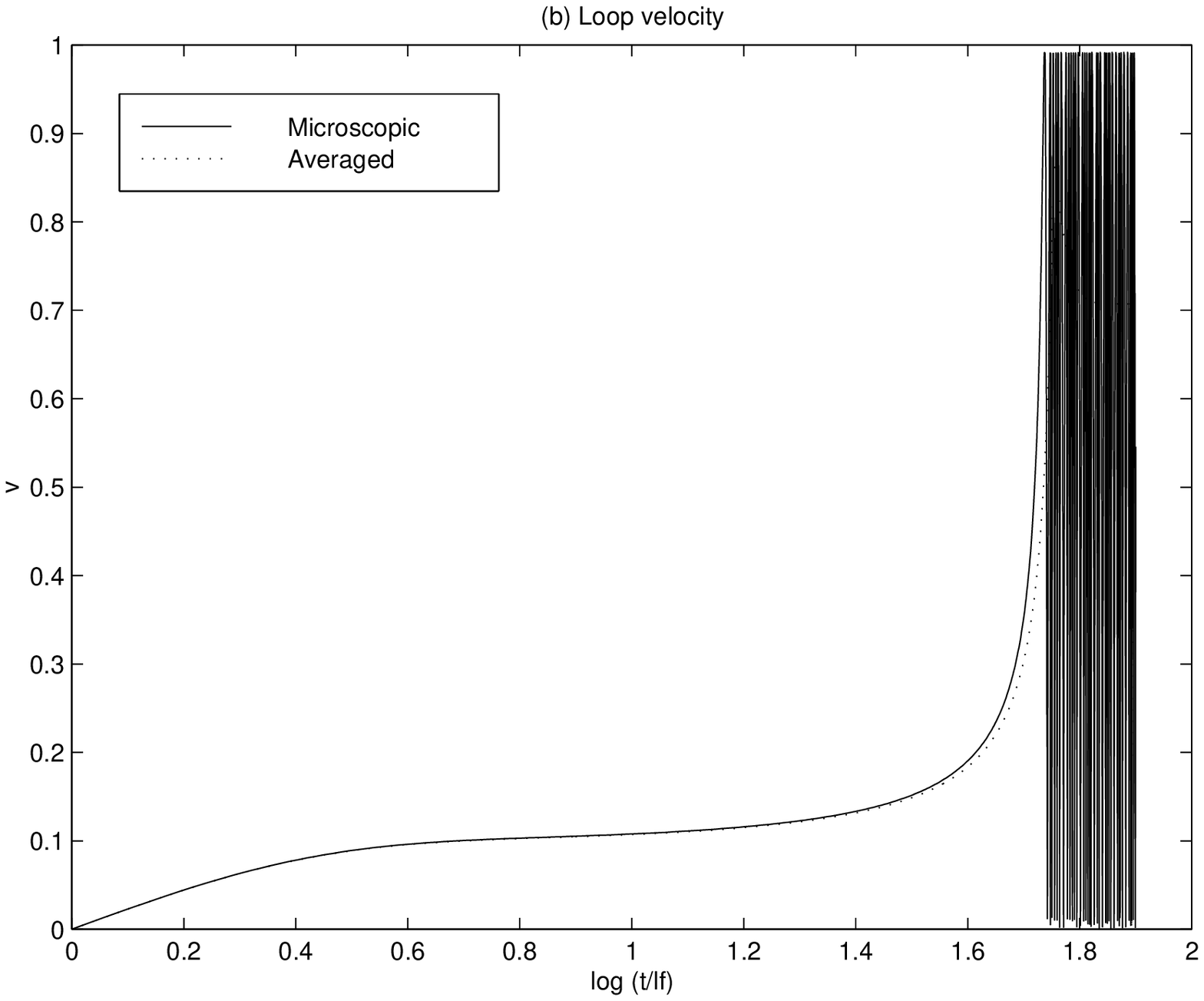}}
\vskip0.2in}
\vbox{\centerline{%
\hskip-3em\epsfxsize=.5\hsize\epsfbox{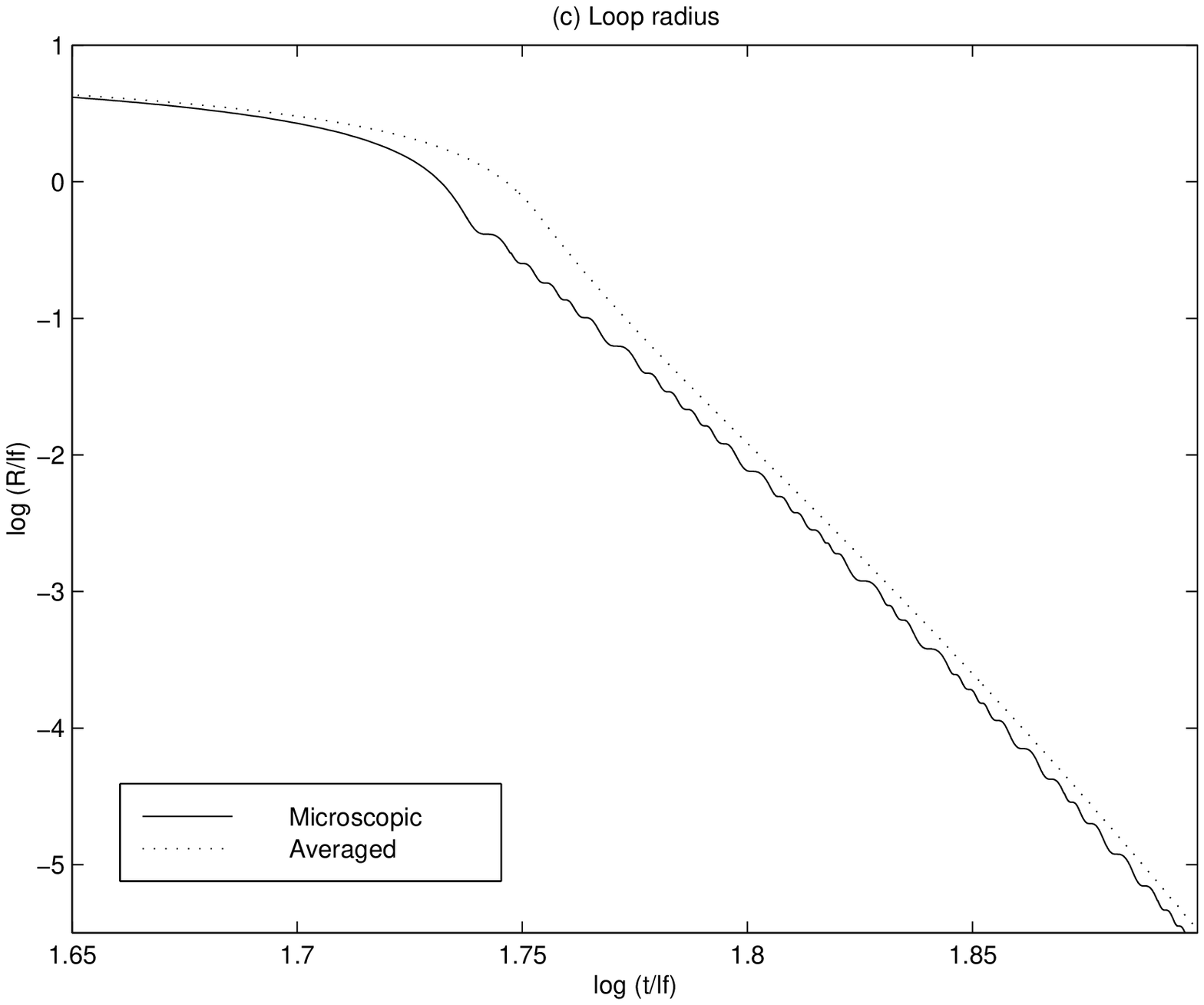}
\hskip3em\epsfxsize=.5\hsize\epsfbox{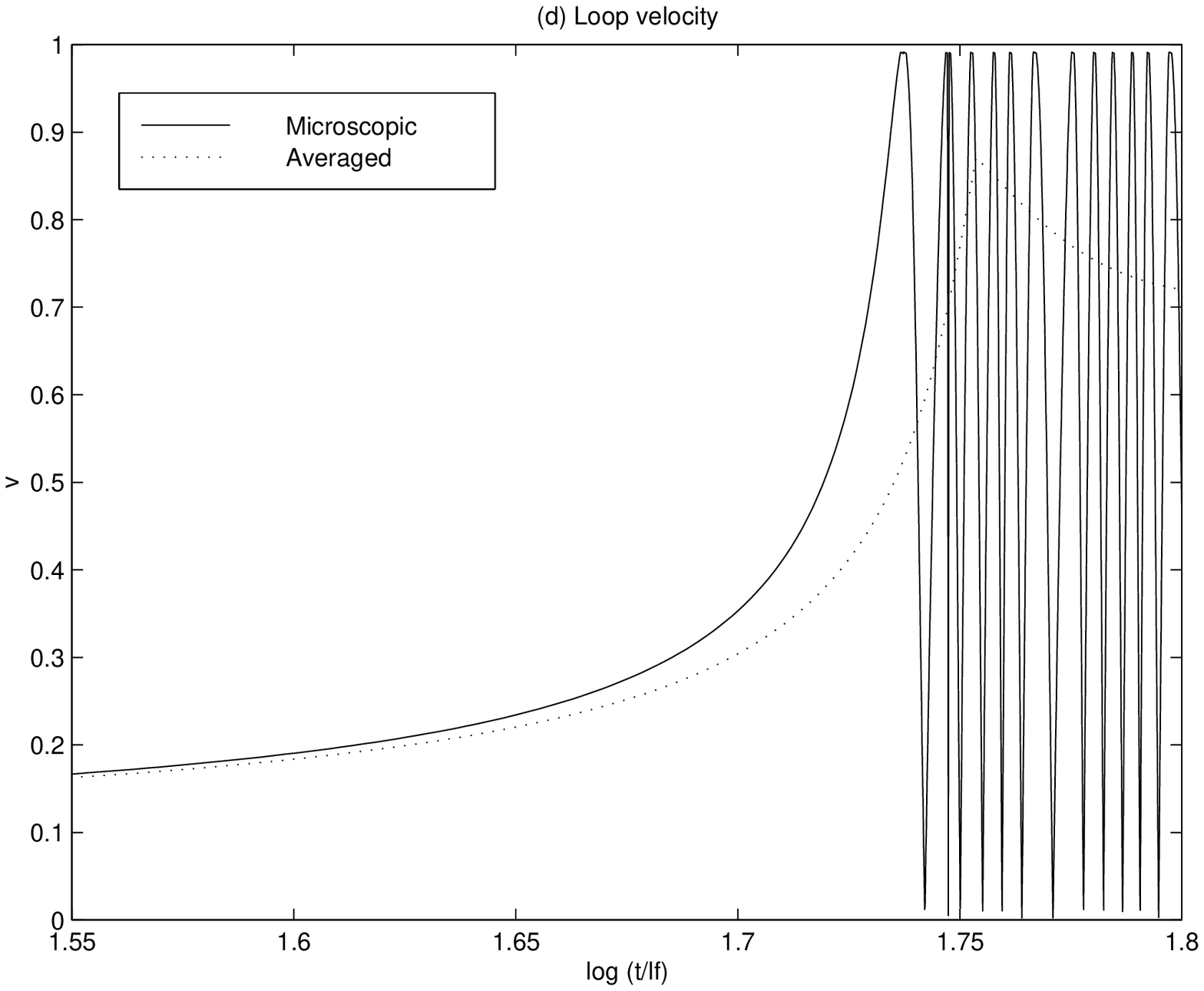}}
\vskip0in}
\caption{Comparing the `microscopic'
(solid lines) and `averaged' (dashed) evolution equations
for a circular loop in a condensed-matter context.
Length and time are in units of $\ell_{\rm f}$, and the time
axis is with a logarithmic scale. Plot (a) depicts the
log of the (rescaled) radius, while (b) depicts the
loop velocity (in units of $c$).
The lower graphs are close-ups of the
upper ones.}
\label{fig32}
\end{figure}

\begin{figure}
\vskip-1in
\vbox{\centerline{%
\hskip-3em\epsfxsize=.5\hsize\epsfbox{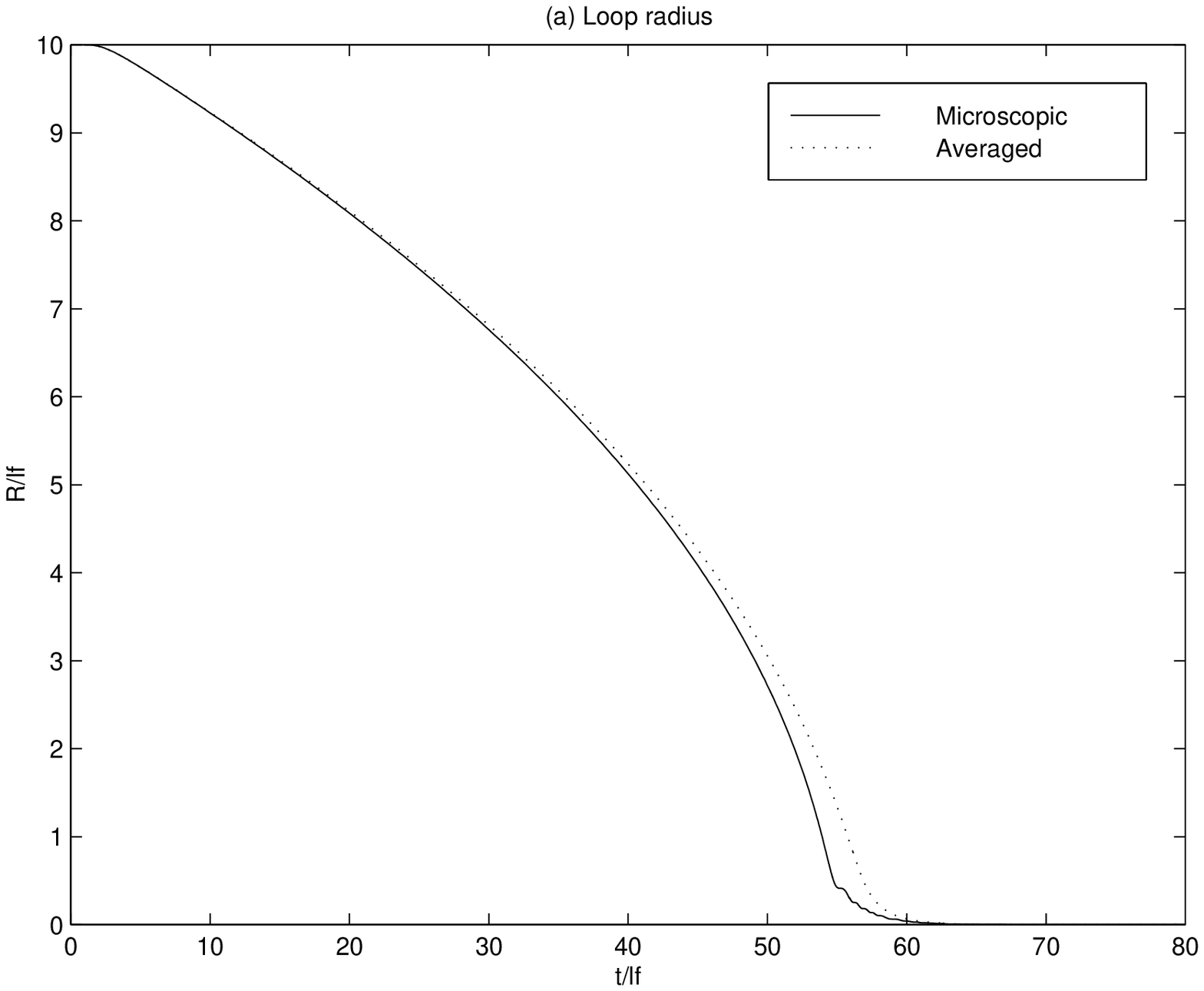}
\hskip3em\epsfxsize=.5\hsize\epsfbox{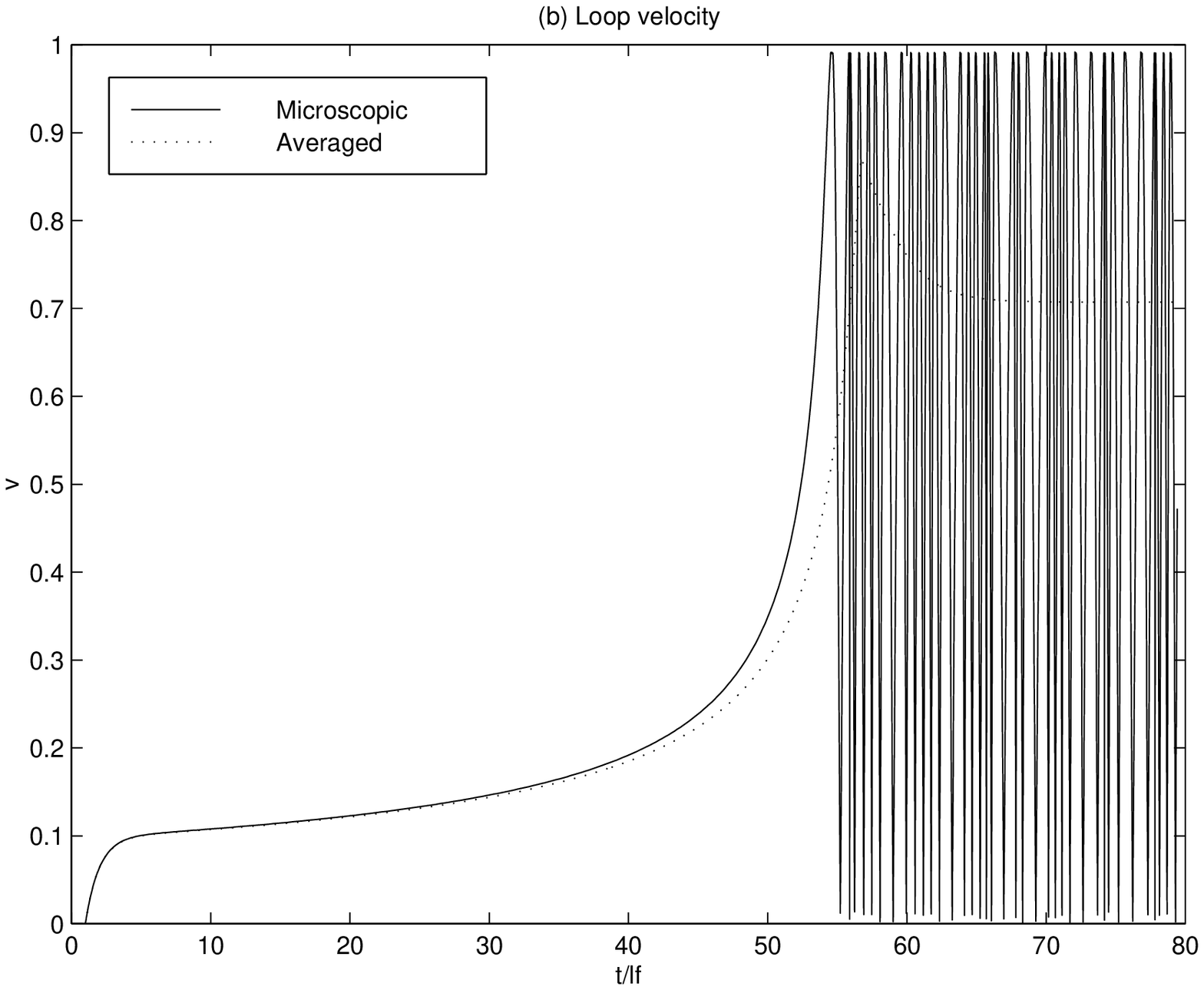}}
\vskip0.2in}
\vbox{\centerline{%
\hskip-3em\epsfxsize=.5\hsize\epsfbox{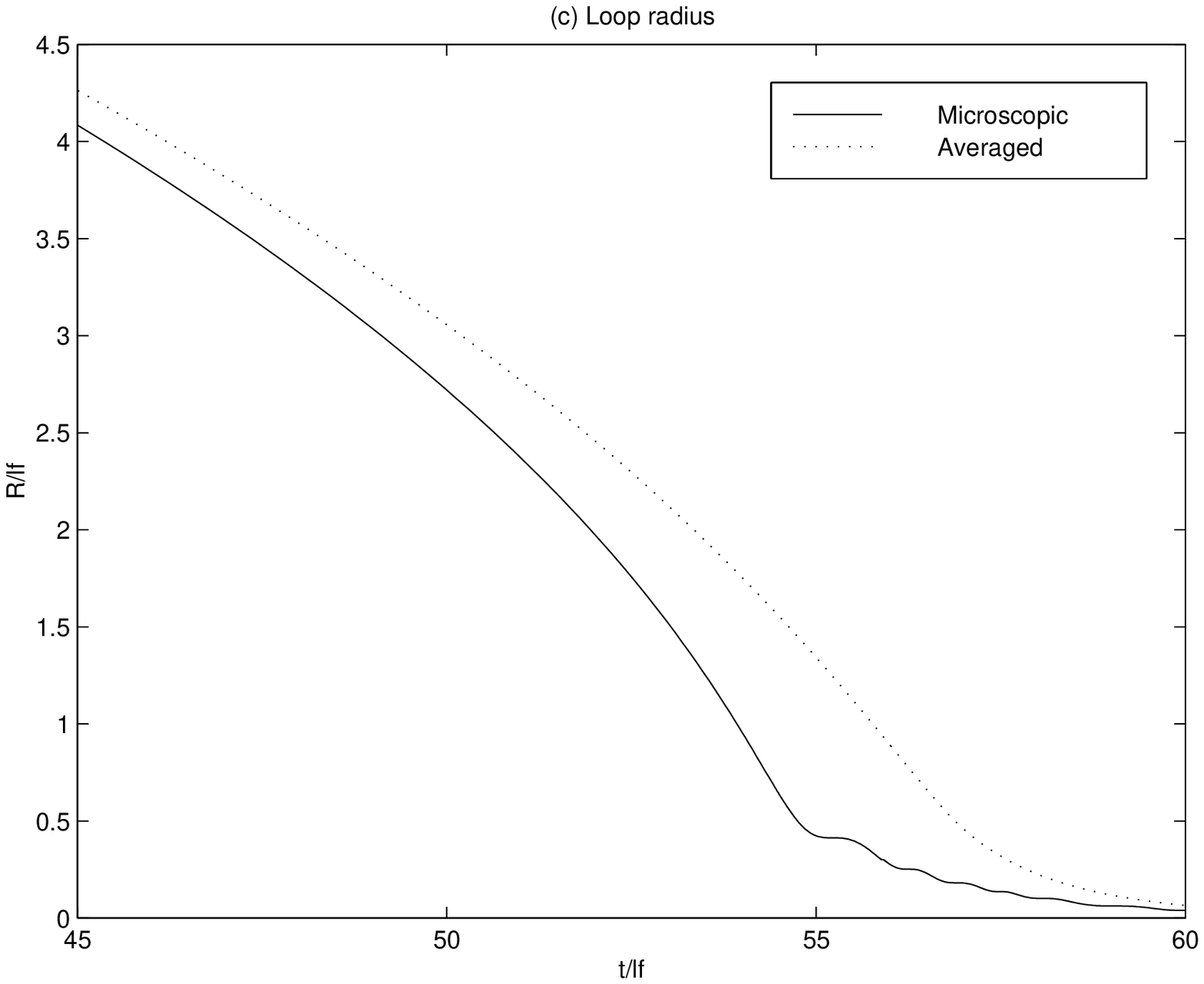}
\hskip3em\epsfxsize=.5\hsize\epsfbox{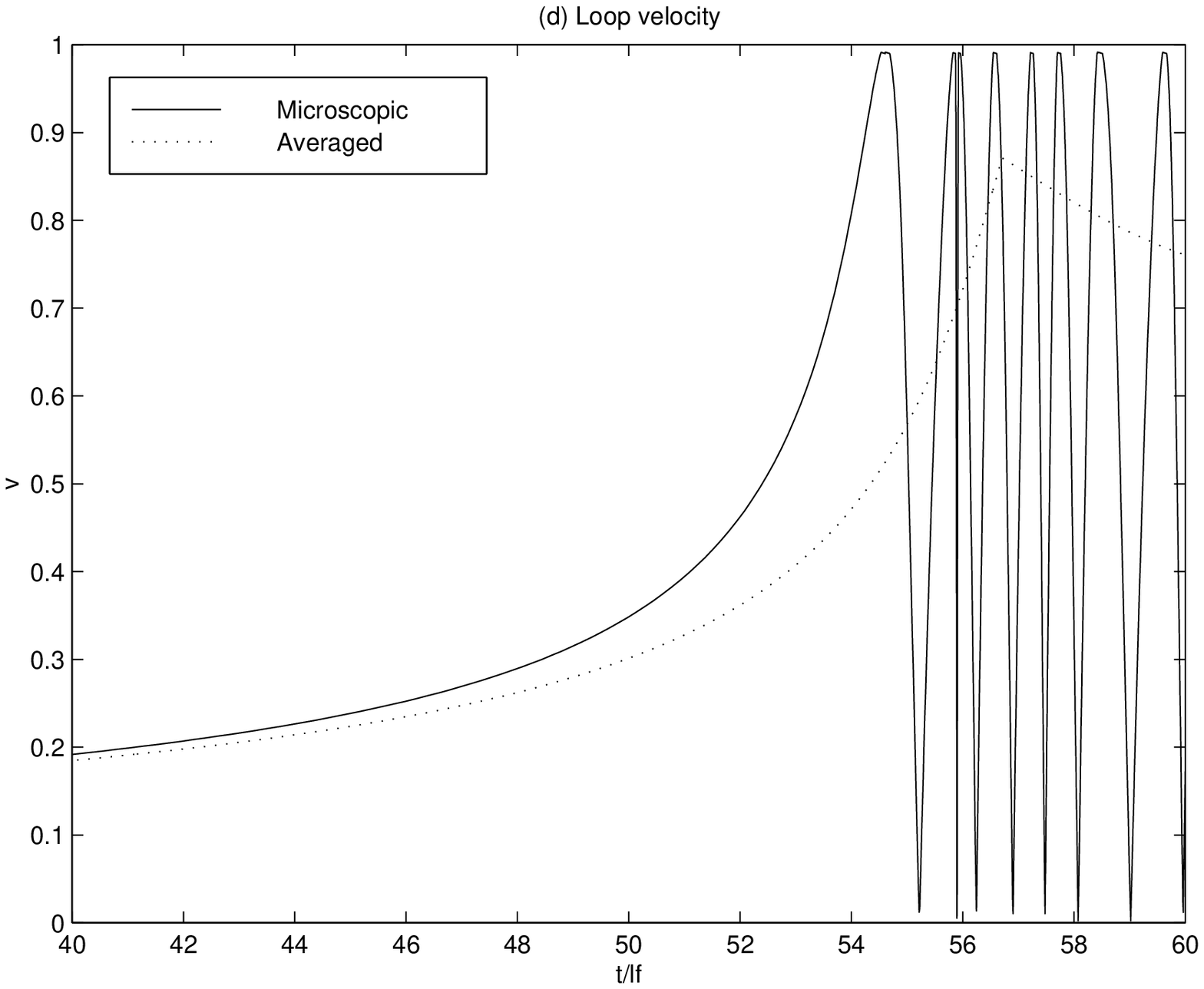}}
\vskip0in}
\caption{Comparing the `microscopic'
(solid lines) and `averaged' (dashed) evolution equations
for a circular loop in a condensed-matter context.
Length and time are in units of
$\ell_{\rm f}$, and the time
axis is with a linear scale. Plot (a) depicts the log
of the (rescaled) radius, while (b) depicts the
loop velocity (in units of $c$).
The lower graphs are close-ups of the
upper ones.}
\label{fig33}
\end{figure}

After numerically comparing the averaged and microscopic
evolution equations, we find that the simplest
possibility,
\begin{equation}
k=\left\{ \begin{array}{ll}
1\, , & \mbox{$\frac{R}{\ell_{\rm f}}>\chi$} \\
\frac{1}{\sqrt2}\frac{R}{\ell_{\rm f}}\, , &
\mbox{$\frac{R}{\ell_{\rm f}}<\chi$}
 \end{array} \right.
\label{fkans}
\end{equation}
(see figure \ref{fig31}) provides the best answer (see
figures \ref{fig32},\ref{fig33}). In particular, this
turns out to be significantly better than assuming
smoother (and slower) transitions between the two
regimes. As can be readily seen, this ansatz provides a
very good fit, considering the lack of parameters
available.

In passing, it is worth pointing out that one can also
easily calculate the loop lifetime\cite{gs}. In the
relativistic regime, the ${\bar R}$ evolution equation
can be written
\begin{equation}
\frac{d{\bar R}}{dt}=-\frac{{\bar R}}{\tau_{\rm f}}
\, , \label{avevrel}
\end{equation}
so we can immediately estimate that the loop will
disappear in a time $t_{dec}\sim2\tau_{\rm f}$
after its first collapse.

Having thus established, in a simple but physically
relevant case, the validity of our generalized
`one-scale model', and in particular of the ansatz for
$k$, we now proceed to apply it to the study of string
evolution in condensed matter contexts.

\section{String evolution at constant temperature}
\subsection{The condensed-matter context}
\subsubsection{Network scaling}
As we already noted, in this case the dynamics is always
dominated by friction. This means that the `correlation
length' $L$ should always be larger than the (constant)
friction length, so we can take $k=1$. Then the evolution
equations can be approximated by

\begin{equation}
2 \frac{dL}{dt}=L\frac{v^2}{\Gamma}+{\tilde c}v
\, , \label{evlcml}
\end{equation}
\begin{equation}
\frac{dv}{dt}=c^2\left(\frac{1}{L}
-\frac{v}{\Gamma}\right)
\, ; \label{evlcmv}
\end{equation}
the friction lengthscale and the string energy per unit
length being respectively
\begin{equation}
\ell_{\rm f}=\left\{ \begin{array}{ll}
s & \mbox{Gauge} \\
s \ln\left(\frac{L}{\delta}\right) & \mbox{Global}
 \end{array} \right.
\label{lfcm}
\end{equation}
and
\begin{equation}
\mu=\left\{ \begin{array}{ll}
T_c^2 & \mbox{Gauge} \\
T_c^2 \ln\left(\frac{L}{\delta}\right) & \mbox{Global}
 \end{array} \right.
\label{mucm}
\end{equation}
where $T_c$ is the temperature at which the strings form
and $s$ is a constant. We then find the following
late-time asymptotic behaviour:
\begin{equation}
L=\sqrt{1+{\tilde c}}\left(\Gamma t\right)^{1/2}
\, ,\label{bhlcm}
\end{equation}
\begin{equation}
v=\frac{\Gamma}{L}\, .\label{bhvcm}
\end{equation}
Note that in both cases the asymptotic behaviour of the
long-string density is
\begin{equation}
\rho_{\infty}=\frac{(k_BT_c)^2}{(1+{\tilde c})\hbar sc^2t}
\, ; \label{lsdcdm}
\end{equation}
the extra logarithmic dependencies in the global case
cancel out. It should be emphasized that in
condensed-matter analyses one does not consider loop
formation, although there is experimental\cite{cdty} and
computational
evidence for them. Our
results show that loop formation can play an important
evolutionary role. The asymptotic ratio of the loop
production and friction terms is a constant, which is
precisely ${\tilde c}$---which in this way acquires a
clearer physical meaning. As expected, increasing
${\tilde c}$ (or including loop losses in the first
place) leads to a lower network scaling density and a
smaller average velocity $v$; furthermore, the approach
to the scaling regime is also faster.

\begin{figure}
\vskip-1in
\vbox{\centerline{%
\hskip-3em\epsfxsize=.5\hsize\epsfbox{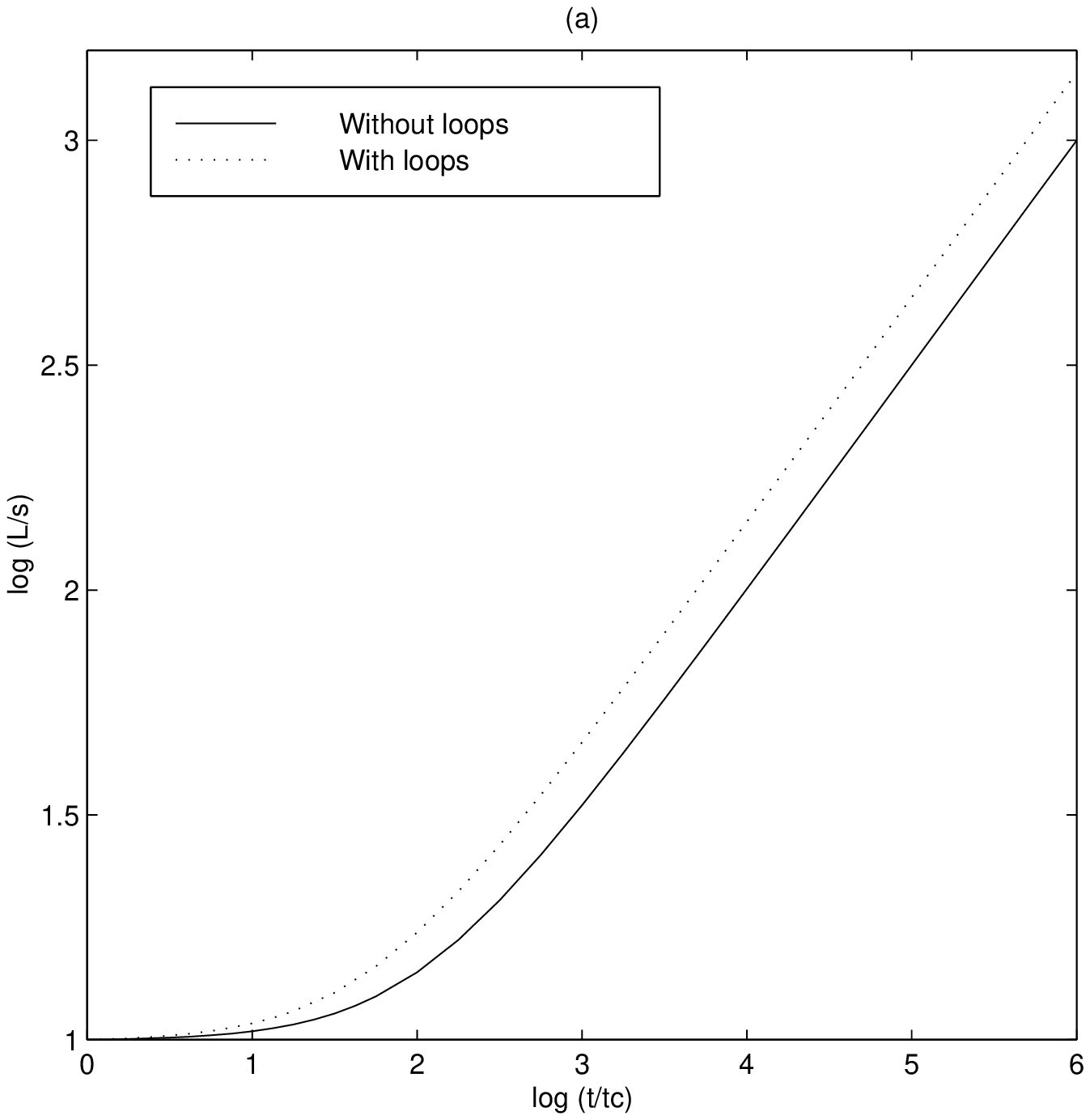}
\hskip3em\epsfxsize=.5\hsize\epsfbox{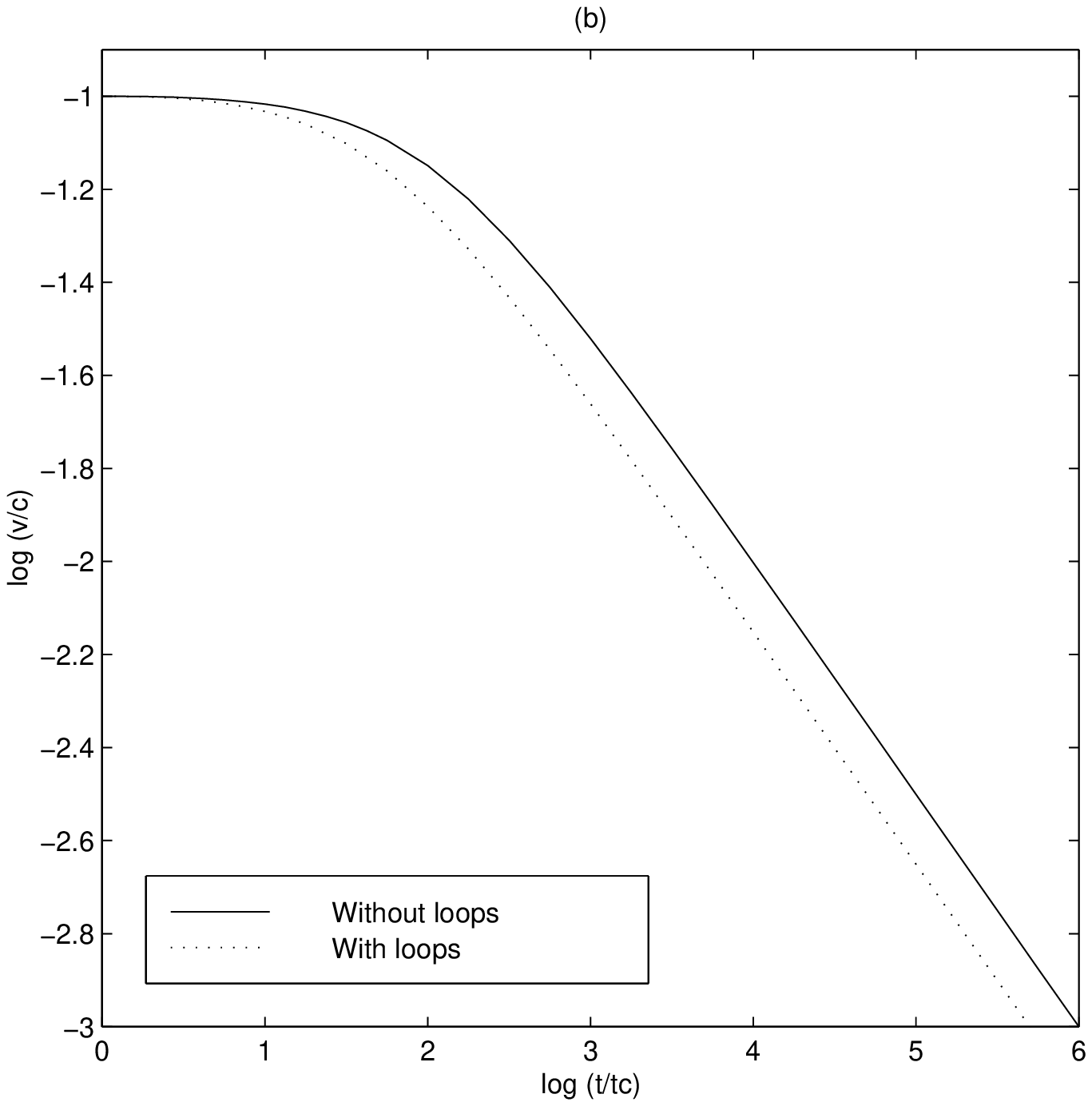}}
\vskip0.2in}
\vbox{\centerline{%
\hskip-3em\epsfxsize=.5\hsize\epsfbox{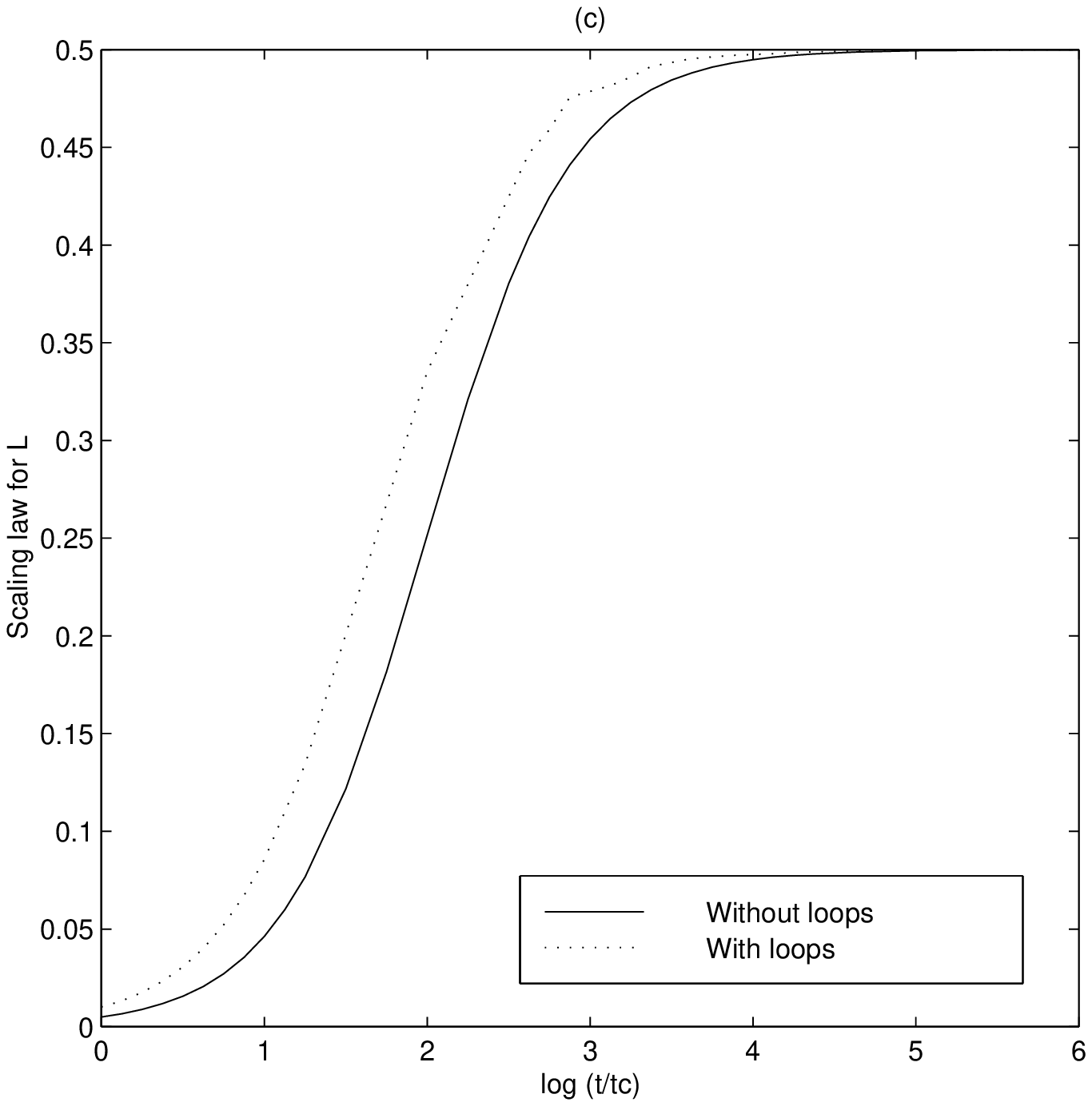}
\hskip3em\epsfxsize=.5\hsize\epsfbox{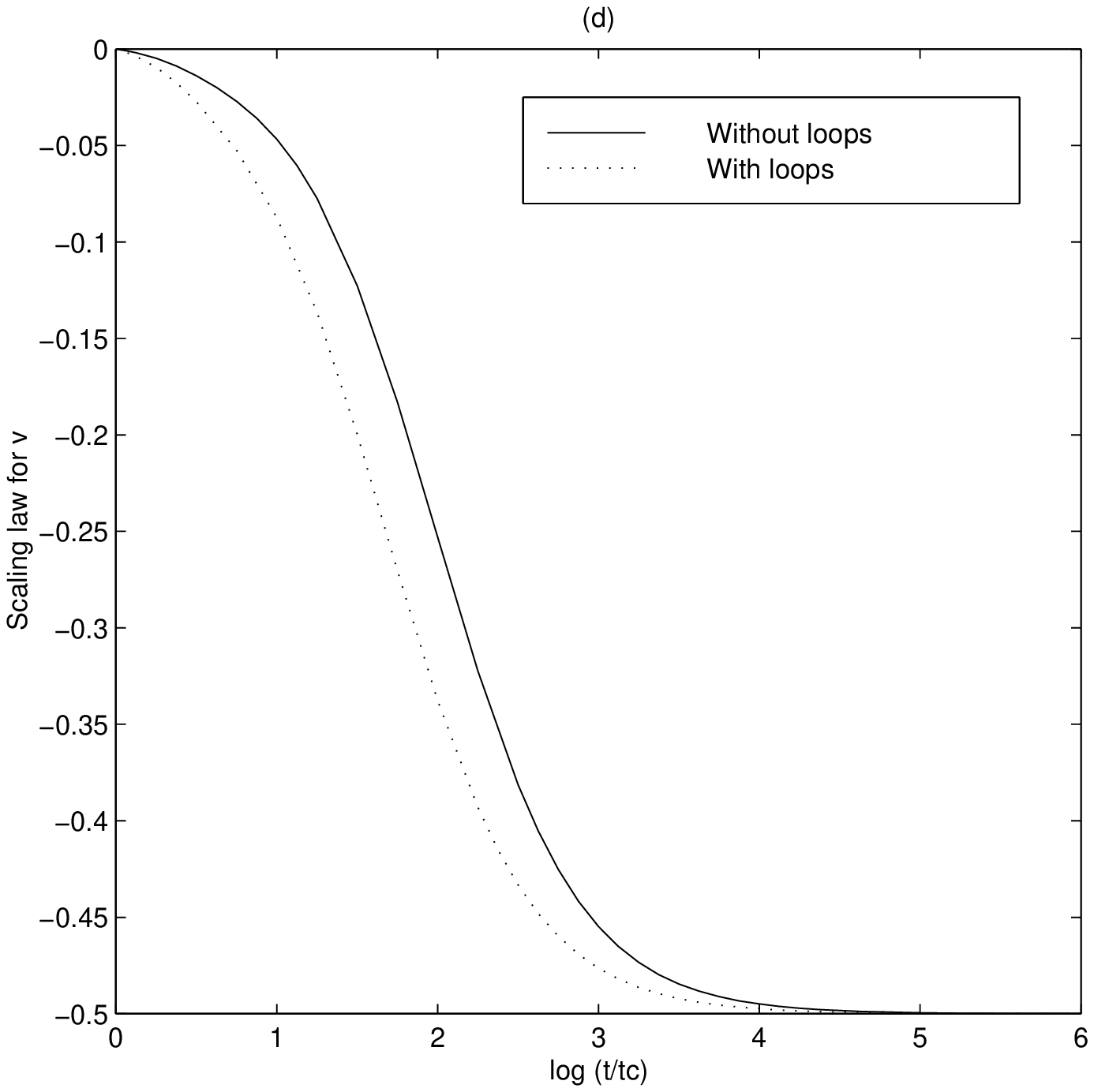}}
\vskip0in}
\caption{The approach to scaling of a
network of gauge strings at constant temperature for
${\tilde c}=0$ (solid lines) and ${\tilde c}=1$
(dotted lines); $k=1$ and the initial conditions are
$0.1 L_i=ct_i=s$, $v_i=0.1\, c$. The
plots correspond to the evolution of $L/s$ (a)
and $v$ (in units of $c$) (b) and the exponents of the power-law
dependence of $L$ (c) and
$v$ (d). The horizontal axis is labeled in orders
of magnitude in
time form the moment of string formation; in (a)
an (b), the `y' axis is also in a
logarithmic scale.}
\label{fig41}
\end{figure}

Figure \ref{fig41} illustrates the relaxation to scaling
of a network of gauge strings for a particular set of
initial conditions, with and without loop production.
The differences between the two cases are clearly
visible. Notice that in the gauge case the temperature
only enters the scaling solution in the prefactor $s$;
hence the plots of figure \ref{fig41} (where the
dimensionless variable $L/s$ is used) are `universal
curves', that is, hold for all relevant temperatures.

\begin{figure}
\vskip-1in
\vbox{\centerline{%
\hskip-3em\epsfxsize=.5\hsize\epsfbox{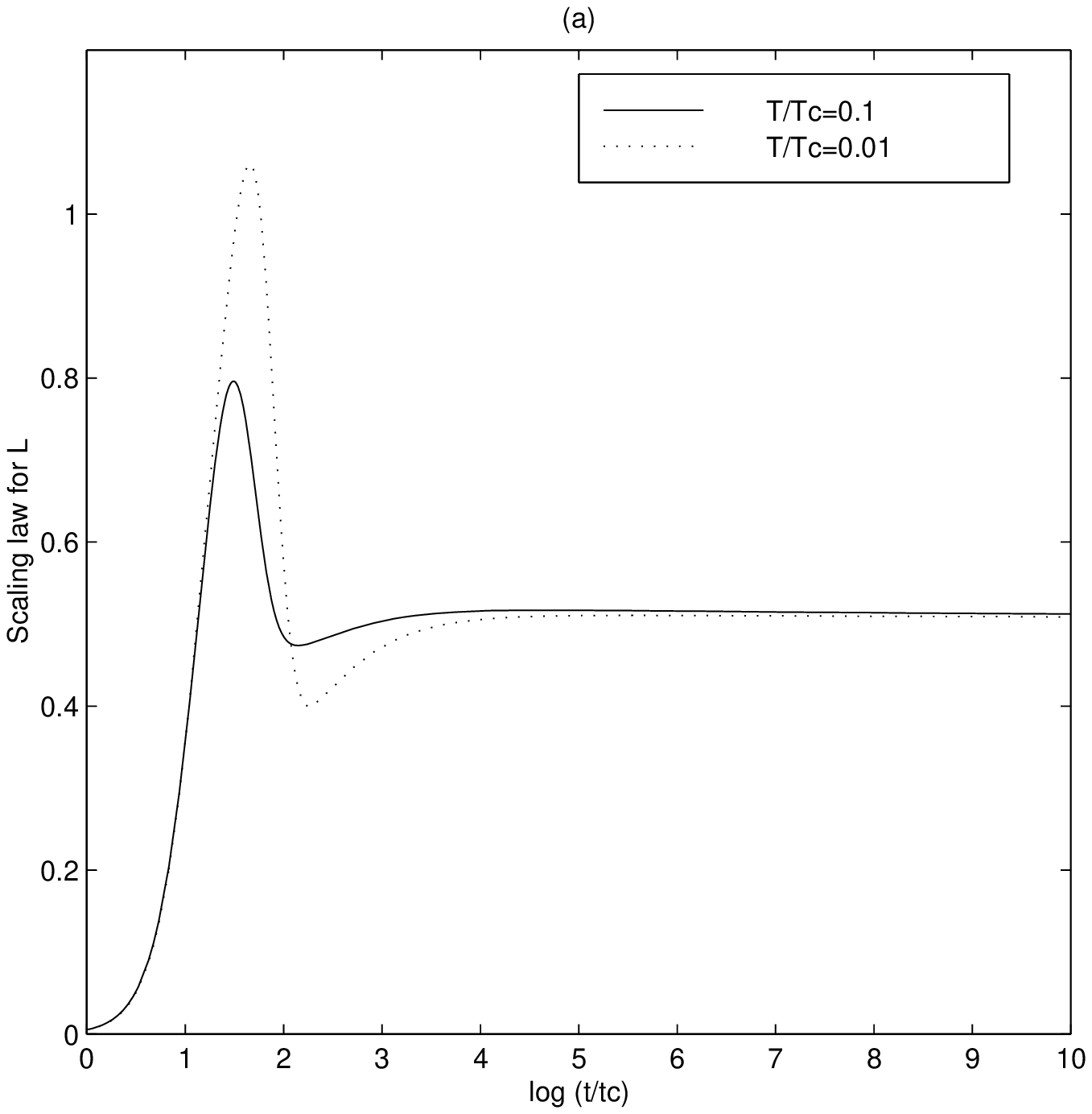}
\hskip3em\epsfxsize=.5\hsize\epsfbox{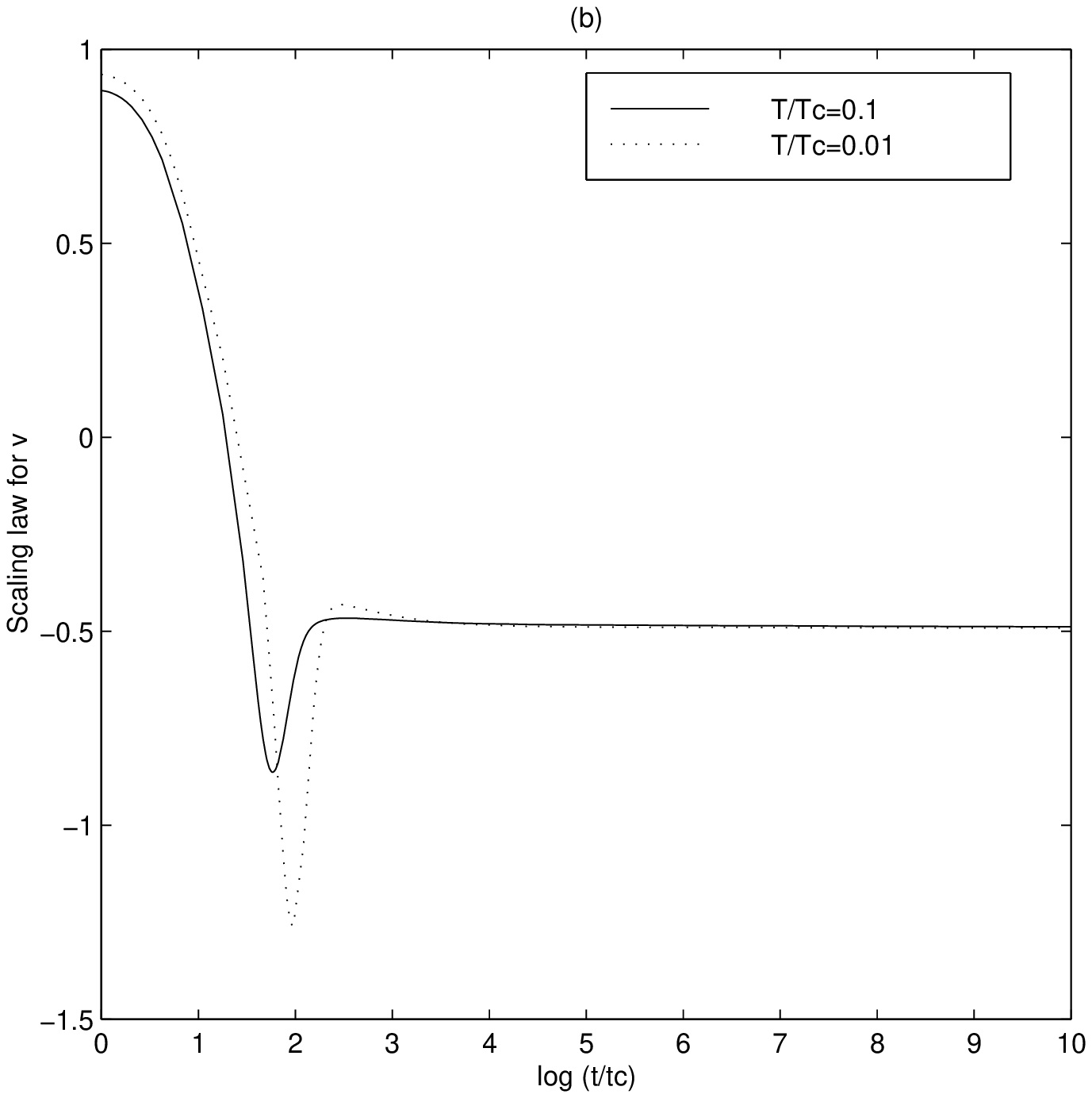}}
\vskip0.2in}
\vbox{\centerline{%
\hskip-3em\epsfxsize=.5\hsize\epsfbox{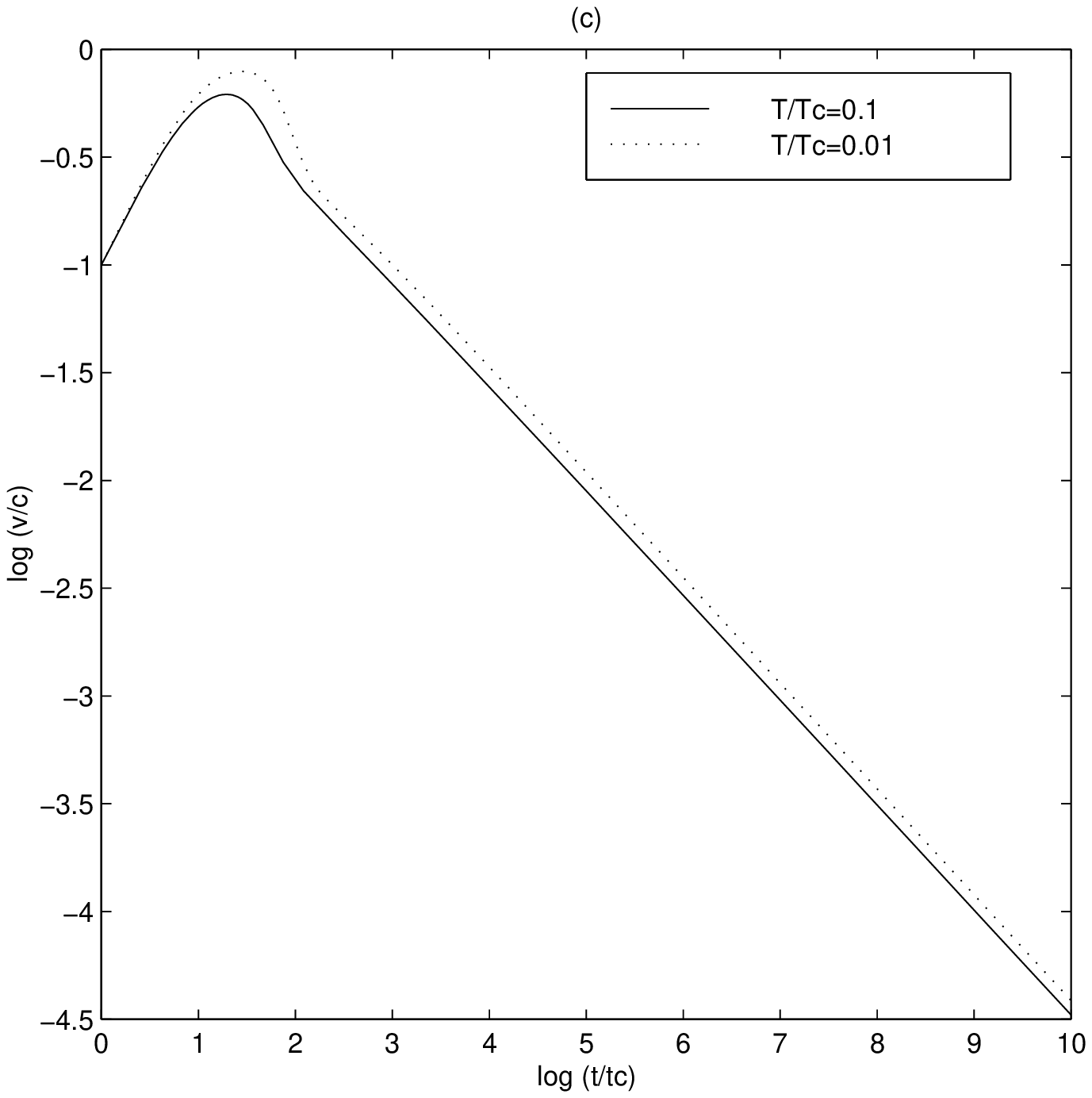}
\hskip3em\epsfxsize=.5\hsize\epsfbox{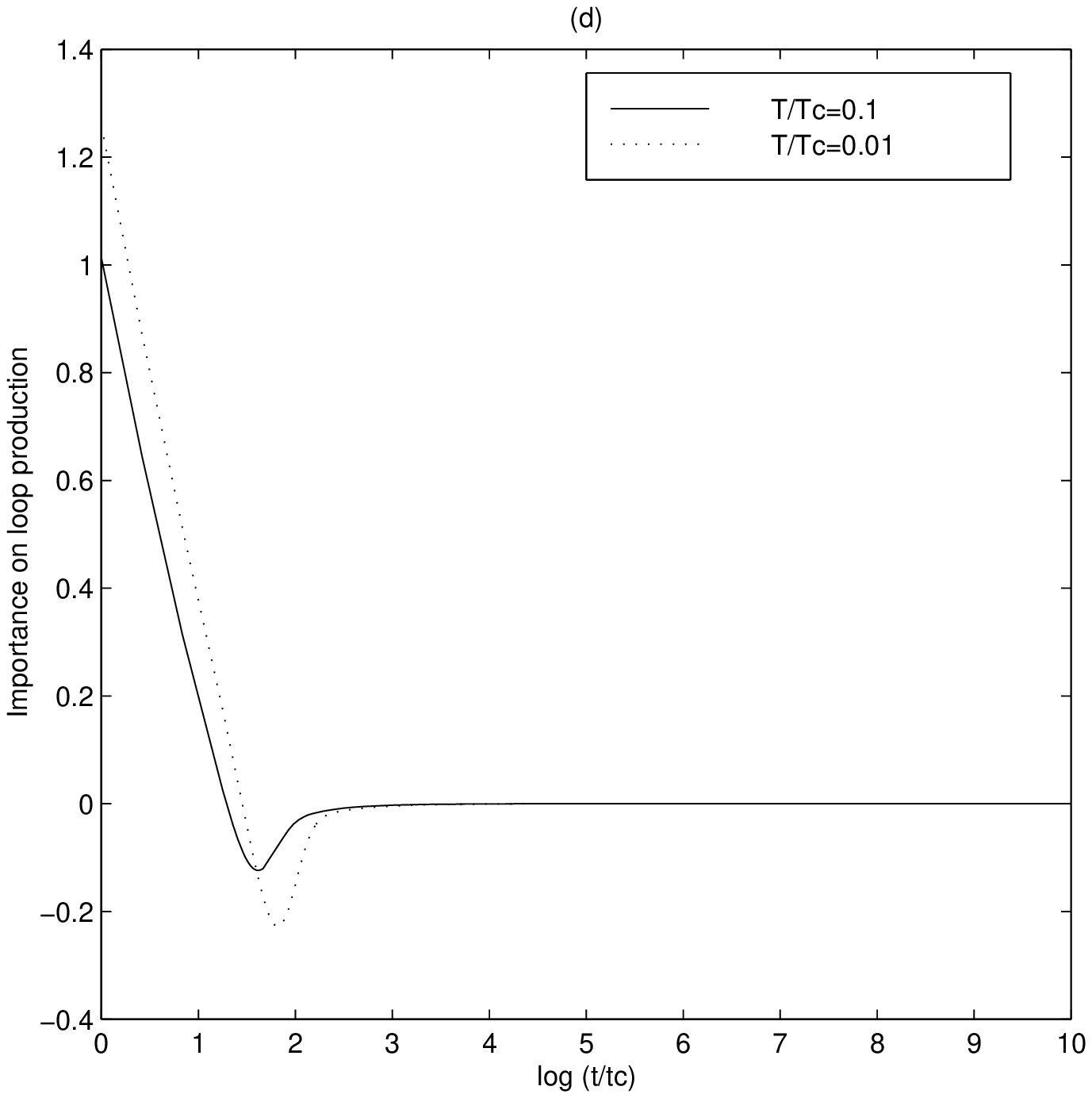}}
\vskip0in}
\caption{The approach to scaling of a network
of global strings at constant temperature for
$T/T_c=0.1$ (solid lines) and $T/T_c=0.01$ (dotted
lines); ${\tilde c}=1$, $k=1$ and the initial
conditions are $0.1 L_i=ct_i=s$, $v_i=0.1\, c$. The
plots correspond to the evolution of the exponents
of the power-law dependence of $L$ (a) and $v$ (b),
$v$ itself (in units of $c$) (c) and the ratio
of the loop formation and
friction terms (d). The horizontal axis is labeled
in orders of magnitude in time from the moment of
string formation; in (c) an (d), the `y' axis is
also in a logarithmic scale.}
\label{fig42}
\end{figure}

In the global case, the logarithmic dependence of the
friction lengthscale gives rise to an additional
logarithmic dependence of the scaling solution on
temperature. One can define $d$ via
$L/\delta\equiv L/sd$; in the Goldstone model we then
have
\begin{equation}
d^{-1}(T)=\sqrt{\frac{\lambda}{6}} \left(\frac{T_c}{T}
\right)^3 \ln^2\left(\sqrt{\frac{\lambda}{6}}
\frac{T}{T_c}\right)
\, . \label{ddt}
\end{equation}
The approach to scaling of a network of global strings
is shown in figure \ref{fig42} for two different
temperatures. Note the enhancement of loop production at
the early stages, since the string velocity is high;
correspondingly, there is a fast growth of the
correlation length. Comparing with the gauge case
(see figure \ref{fig41}), one finds that the effect of
the extra logarithmic terms is significant for at least
three orders of magnitude in time.

\subsubsection{Loop populations}
Now let us consider the loop populations. In the
traditional (that is, `static') approach one would
simply calculate the loop distribution
$\rho_\ell (\ell,t)$ in the scaling regime---which we
assume starts at some time $t_s$. The corresponding
evolution equation is then
\begin{equation}
\frac{d\rho_{\ell}}{dt}+
\frac{v^2_{\ell}}{\Gamma}\rho_{\ell}=g\mu\frac{v\ell}{
L^5}w\left(\frac{\ell}{L}\right)
\, . \label{lpdcm}
\end{equation}
Since the loop size and velocity are initially
(that is, while the loops are overcritically damped)
slowly-varying quantities, this approach will only be
accurate at the early stages of evolution of each
lengthscale $\ell$. Afterwards $\ell$ and $v$ will vary
quickly (the averaged $v$ will be a constant later), and
in this situation this approach is no longer useful.
Note that since friction is dominating the dynamics,
irregularities are quickly erased and all loops become
circular---hence the discussion of section 3 is
particularly relevant here.

Consistent with what we found there, we will use the
following simplifying ansatz
\begin{equation}
\frac{v_\ell (R,t)}{c}=\left\{ \begin{array}{ll}
\frac{\ell_{\rm f}}{R} & \mbox{$\frac{R}{\ell_{\rm f}}
\geq\sqrt{2}$} \\ \frac{1}{\sqrt2} &
\mbox{$\frac{R}{\ell_{\rm f}}\leq\sqrt{2}$}
 \end{array} \right.
\label{vloopans}
\end{equation}
(where $\ell=2\pi R$); naturally in this approach only
the first case is relevant. Furthermore, the loop
production parameter $\alpha$ (defined is (\ref{inls}))
should be a constant (we are considering a scaling
regime) of order unity.

After some algebra, we find the following solution
\begin{equation}
\rho_{\ell}(\ell,t)=\left\{ \begin{array}{ll}
\frac{2g{\tilde c}\alpha^2\mu}{1+{\tilde
c}}\ell^{-3}\exp{\left[-4\pi^2c\ell_{\rm f}t\left(\frac{1}
{\ell^2}-\frac{1}{\alpha^2L^2}\right)\right]}
& \mbox{$\alpha L_s<\ell <\alpha L$} \\
0 & \mbox{otherwise}
 \end{array} \right.
\label{lpldelta}
\end{equation}
where we have neglected any loops present at time $t_s$.
Note that in the global string case these expressions
are only approximate, since there are additional
logarithmic dependencies in the energy per unit length
$\mu$ and the friction length $\ell_{\rm f}$.

However, we can do more than that. To begin with, we can
determine the loop lifetime. During most of it the loop
will have a length greater than the friction lenghtscale,
so we can approximate their velocity by (\ref{vloopans}).
Then its length will vary according to
\begin{equation}
\frac{d\ell}{dt}\approx-4\pi^2
\frac{\Gamma}{\ell} \, , \label{estimlooplife}
\end{equation}
so its lifetime is
\begin{equation}
\frac{\tau (\ell)}{\tau_{\rm f}}=\frac{1}{8\pi^2}\left(
\frac{\ell}{\ell_{\rm f}}\right)^2
\, . \label{teollifecdm}
\end{equation}

\begin{figure}
\vskip-0.1in
\vbox{\centerline{
\epsfxsize=.7\hsize\epsfbox{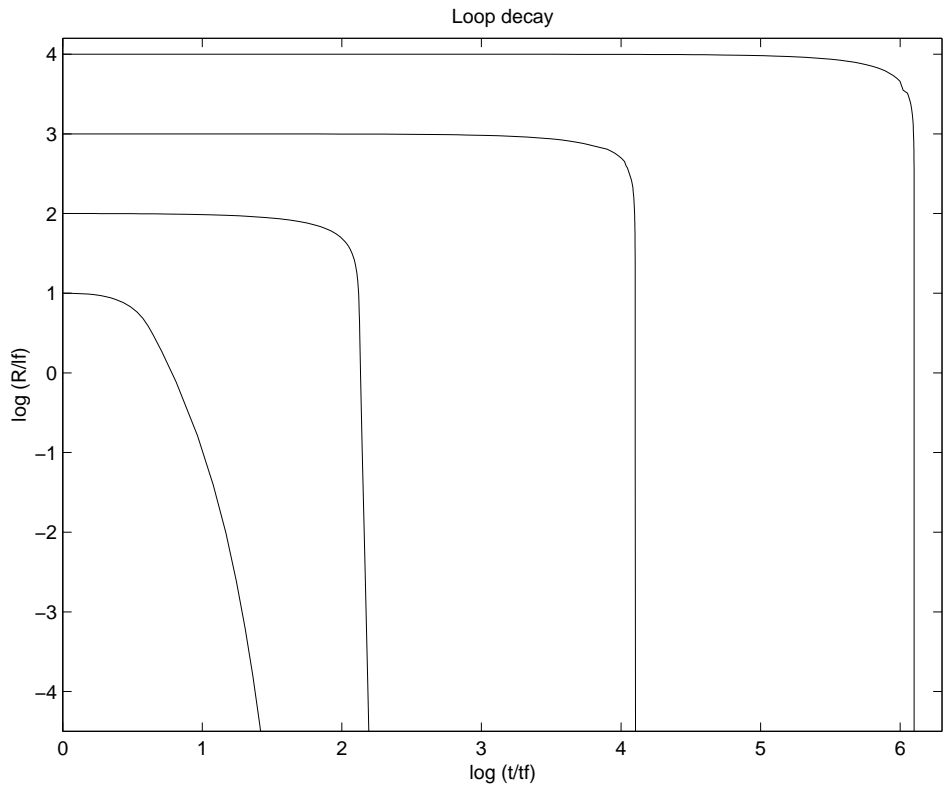}}
\vskip.5in}
\caption{The evolution of loops of different
sizes relative to the friction lengthscale. For
large enough loops, the lifetime is
proportional to the square of the length.}
\label{fig43}
\end{figure}

Of course a more accurate result can be obtained
numerically (see figure \ref{fig43}); for large
enough loops we find
\begin{equation}
\frac{\tau (\ell)}{\tau_{\rm f}}=\iota\left(\frac{\ell}
{\ell_{\rm f}}\right)^2 \, , \, \iota\sim0.0126
\, , \label{llifecdm}
\end{equation}
showing our analytical estimate to be correct to within
a fraction of one percent.

On the other hand, we can simply use (\ref{ratio}) to
determine the ratio of the energy densities in the form
of loops and long strings. In this case, taking
$\alpha\sim1$, it simplifies to
\begin{equation}
\varrho (t)=\frac{\rho_o(t)}{\rho_{\infty}(t)}=
{\tilde c}L^2(t)\int_{t_c}^{t}\frac{v_\infty (t')}{
L^4(t')}\ell(t,t')dt'
\, . \label{ratiod}
\end{equation}

\begin{figure}
\vbox{\centerline{
\epsfxsize=.5\hsize\epsfbox{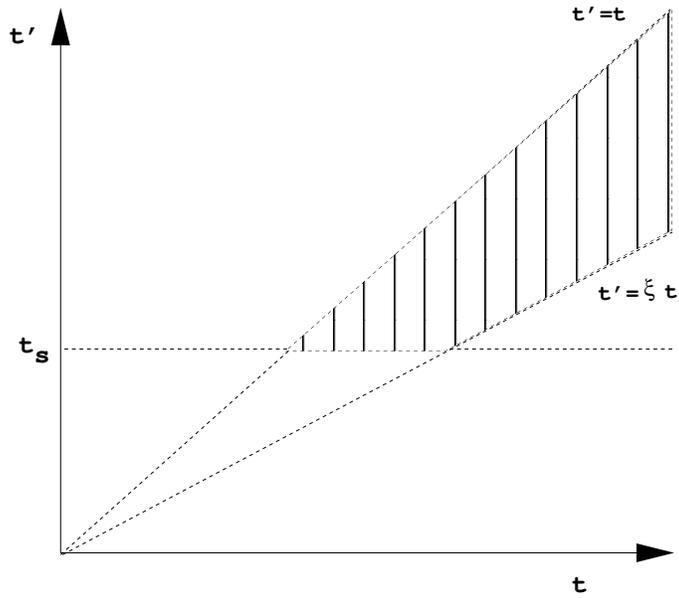}}
\vskip.2in}
\caption{The interval in t' giving a non-zero loop
length contribution to the integral
(\protect\ref{ratiod}) for different
times t.}
\label{fig44}
\end{figure}

In fact, an approximate analytic solution for $\varrho$
can be found in the scaling regime. We will assume that
such a regime starts at a time $t_s$ and neglect any
existing loops at $t_s$. The contribution of each loop
to the loop density will only be significant while it
overdamped. In this regime we can assume that its length
is approximately constant,
$\ell(t,t')\sim\ell(t')=L(t')$ (hence we will be
deriving an overestimate of $\varrho$). On the other
hand, the end of this regime approximately coincides with
the moment of loop decay, which according to
(\ref{llifecdm}) for a loop formed at time $t'$
happens at
\begin{equation}
t\equiv\xi^{-1}t'=\left[1+\iota(1+{\tilde c})\right]t'
\, , \label{estimativa}
\end{equation}
for a loop formed at time $t'$. Then we have
\begin{equation}
\ell(t,t')=\left\{ \begin{array}{ll}
L(t') & \mbox{$t'\le t\le \xi t'$} \\
0 & \mbox{otherwise}
 \end{array} \right.
\label{lttpcdmcase}
\end{equation}
and there are two corresponding cases to consider for
the integral (\ref{ratiod}) (see figure \ref{fig44}),
according to whether or not loops have decayed at the
time in question,
\begin{equation}
\varrho (t)={\tilde c}L^2(t)\int_{f(t)}^{t}
\frac{v_\infty (t')}{L^3(t')}dt'
\, , \label{ratiotres}
\end{equation}
where
\begin{equation}
f(t)=\left\{ \begin{array}{ll}
t_s & \mbox{$t\le\xi^{-1}t_s$} \\
\xi t & \mbox{$t\ge\xi^{-1}t_s$}
 \end{array} \right.
\label{fdetecondmat}
\end{equation}
We then obtain
\begin{equation}
\varrho_{scaling}=\left\{ \begin{array}{ll}
\frac{{\tilde c}}{1+{\tilde c}}\left(\frac{t}{t_s}
-1\right) & \mbox{$t\le\xi^{-1}t_s$} \\
\iota{\tilde c} & \mbox{$t\ge\xi^{-1}t_s$}
 \end{array} \right.
\label{ratioquatro}
\end{equation}

\begin{figure}
\vskip-0.1in
\vbox{\centerline{
\epsfxsize=0.7\hsize\epsfbox{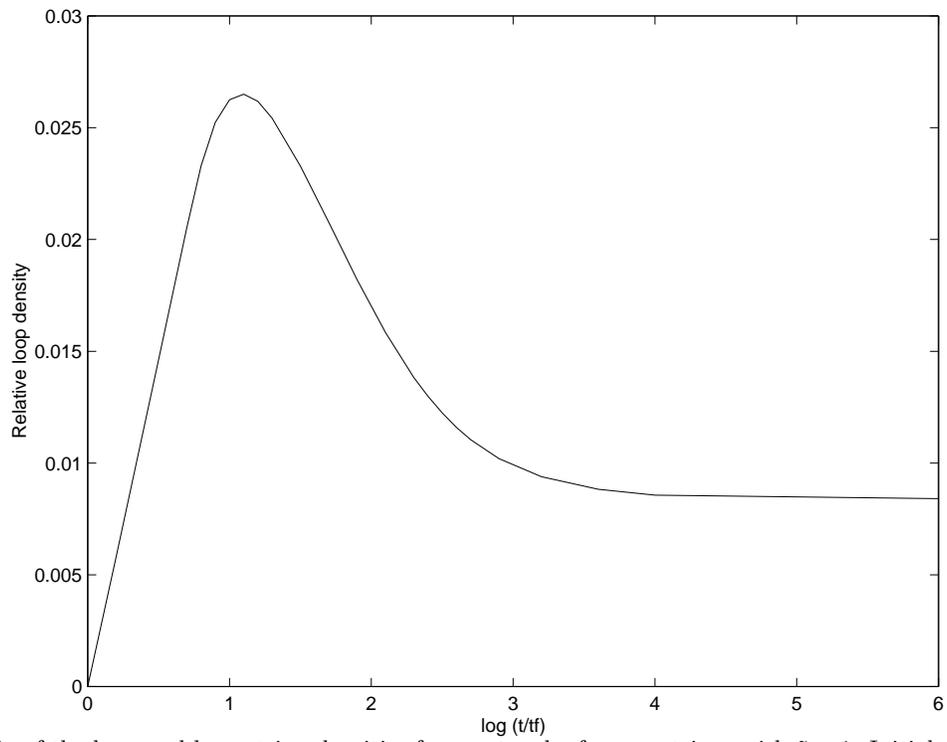}}
\vskip0in}
\caption{The ratio of the loop and long string densities
for a network of gauge strings with ${\tilde c}=1$.
Initial conditions are as in the corresponding case of
figure \protect\ref{fig41}}
\label{fig45}
\end{figure}

This is easily confirmed numerically. Figure \ref{fig45}
displays the evolution of the ratio of the loop and long
string densities corresponding to the gauge string
network of figure \ref{fig41}. In this case
${\tilde c}=1$, so our analytical estimate is
$\varrho_{scaling}=0.0126$, while numerically we find
$\varrho_{scaling}\sim0.0089$. Notice that with our
initial conditions the ratio of the energy densities in
loops and long strings is larger while the network is
approaching scaling than in the scaling regime. This is
simply due to the fact that loops with initial sizes not
much larger than the friction lengthscale live longer
than predicted by (\ref{llifecdm}) (as can be seen in
figure \ref{fig43}).

Naturally, since we can find out how each loop evolves,
it is always possible to find out how many loops there
are at any time on any given length interval. Note that
in this context reconnections and self-intersections
should be negligible, at least when the loops are
large compared to $\ell_{\rm f}$ (because they are
non-relativistic)---which is when their contribution to
the loop density is significant.

Furthermore, using the results from our discussion in
section 2C, we easily find that the fraction of the
energy density in long strings at time $t$ that is
converted into loops in the time interval
$\Delta t=\tau_{\rm f}$ is asymptotically
\begin{equation}
f_{\tau_{\rm f}}=\frac{{\tilde c}}{1+{\tilde c}}
\frac{\tau_{\rm f}}{t}
\, ; \label{fcdmnf}
\end{equation}
note that this is a very small number. Furthermore, note
that as we increase ${\tilde c}$ there is a limiting
fraction. Also, the number of loops formed in a time
interval $\Delta t=\tau_{\rm f}$
and volume $\ell_{\rm f}^3$ is
\begin{equation}
n_{\tau_{\rm f}}=\frac{{\tilde c}}{(1+{\tilde c})^{3/2}}
\left(\frac{\tau_{\rm f}}{t}\right)^{5/2}
\, ; \label{fcdmnn}
\end{equation}
in this case, $n_{\tau_{\rm f}}$ goes to zero as ${\tilde c}$
grows---this is because $L$ and hence the loop size at
formation grow with ${\tilde c}$.

\subsubsection{Discussion}
The $t^{1/2}$ scaling law for the characteristic
lengthscale is a well-known result in the theory of
phase ordering (that is, the growth of `order'---as measured
by some correlation length---by domain coarsening when a
system is `quenched' from a homogeneous phase into a
broken-symmetry phase) with a non-conserved order
parameter. In this context it is usually called the
Lifshitz-Allen-Cahn\cite{lac} growth law, and it is
widely supported by simulations and
experiment\cite{mondello}
(see also ref. \onlinecite{ab} for a recent review).
In the usual approach, one sets up a continuum
description in terms of a coarse-grained order parameter
$\phi$ and then assumes a {\it scaling hypothesis}, that
is, that at late times there is a single lengthscale
such that the domain structure is time-independent (in a
statistical sense) when all lengths are rescaled by it.
The growth law is usually derived by studying the
dynamics of the defects in $\phi$ (see, for example,
Section 3 in Bray's review\cite{ab}). A recent
alternative approach\cite{br}
proceeds instead
by comparing the global rate of energy change due to the
energy dissipation to the local evolution of the order
parameter; with the scaling hypothesis, the
time-dependence of the lengthscale can be determined
self-consistently.

In particular, the $L\propto t^{1/2}$ law has been
experimentally confirmed for the evolution of a string
network in a nematic liquid crystal (roughly speaking, a
liquid made of rod-like molecules)---eg, see ref.
\onlinecite{cdty}
where, as mentioned, loop formation and decay have
been seen.

The $v\propto L^{-1}\ln L$ scaling law is also known in
hydrodynamical contexts. Furthermore, it has been shown
that it holds for superfluid vortex-rings\cite{dsmag} in
the context of a modified Goldstone model (in a way
described in the previous section). Hence the above
result seems to indicate that a global string network at
constant temperature asymptotically
behaves as if it was made of loops of size $L$.

As we hopefully made clear, the above discussion of the
loop lifetime and density is entirely new in this
context. It is hoped that these results, obtained in a
fairly simple way in this approach will stimulate
condensed matter theoreticians an experimentalists...

Finally, in the case of superfluid vortices, despite the
additional Magnus force term, the evolution equations
also have the form
(\ref{evlcml},\ref{evlcmv}). In this case, however,
the physical meaning of the friction lengthscale is not
clear. Furthermore, it is also not clear how one can
describe the effect of the Magnus force on the evolution
of the network. We hope to address these issues
at a later stage.

Therefore, a model aimed at describing the evolution of
cosmic strings\cite{ms,ms1} can be straightforwardly
applied to rather more `down-to-earth' situations, with
some advantages over previously existing analytical
approaches (in particular, allowing a precise treatment
of loop production). This is rather remarkable,
considering the difference in the energy scales involved.

\subsection{The `relativistic' regime}
As a matter of completeness as well as mathematical
curiosity, we now consider the evolution of a string
network in flat space with a constant friction
lengthscale when the initial conditions are such that
the correlation length is much smaller than the friction
lengthscale. This is therefore the opposite regime to
that usually observed in condensed matter.

The evolution equations will now be
\begin{equation}
2 \frac{dL}{dt}=\frac{v^2}{\Gamma}L
+{\tilde c}v
\, , \label{evlcmkl}
\end{equation}
\begin{equation}
\frac{dv}{dt}=\left(c^2-v^2\right)\left(
\frac{k(L)}{L}-\frac{v}{\Gamma}\right)
\, , \label{evlcmkv}
\end{equation}
where we will assume that our ansatz for $k$ for string
loops also holds for long strings (see ref.
\onlinecite{ms1} for a
more complete discussion of this point), that is
\begin{equation}
k=\left\{ \begin{array}{ll}
1\, , & \mbox{$\frac{L}{\ell_{\rm f}}>\chi$} \\
\frac{1}{\sqrt2}\frac{L}{\ell_{\rm f}}\, , &
\mbox{$\frac{L}{\ell_{\rm f}}<\chi$}
\end{array} \right.
\label{k2ans}
\end{equation}

In the regime where $R\ll\ell_{\rm f}$ the $v$-equation is
independent of $L$, so its particularly
easy to find the scaling regime
\begin{equation}
\frac{L}{\ell_{\rm f}}=\left(\frac{L_o}{\ell_{\rm f}}
+2\sqrt{2}{\tilde c}\right)\exp
\left[\frac{c}{4\ell_{\rm f}}(t-t_0)
\right]-2\sqrt{2}{\tilde c} \, , \label{lfast}
\end{equation}
\begin{equation}
\frac{v}{c}=\frac{1}{\sqrt{2}}
\, . \label{vfast}
\end{equation}

\begin{figure}
\vskip-1in
\vbox{\centerline{%
\hskip-3em\epsfxsize=.5\hsize\epsfbox{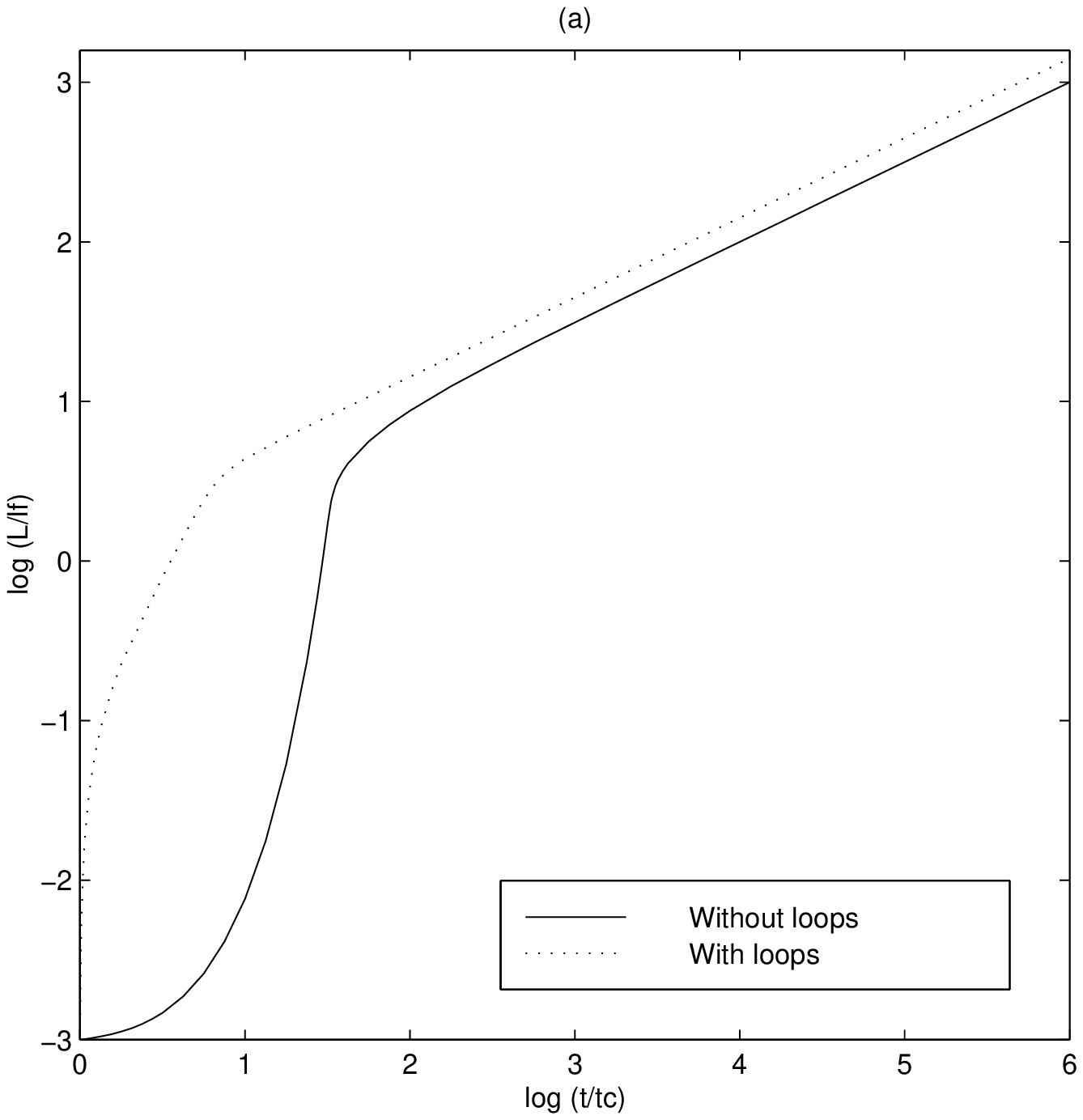}
\hskip3em\epsfxsize=.5\hsize\epsfbox{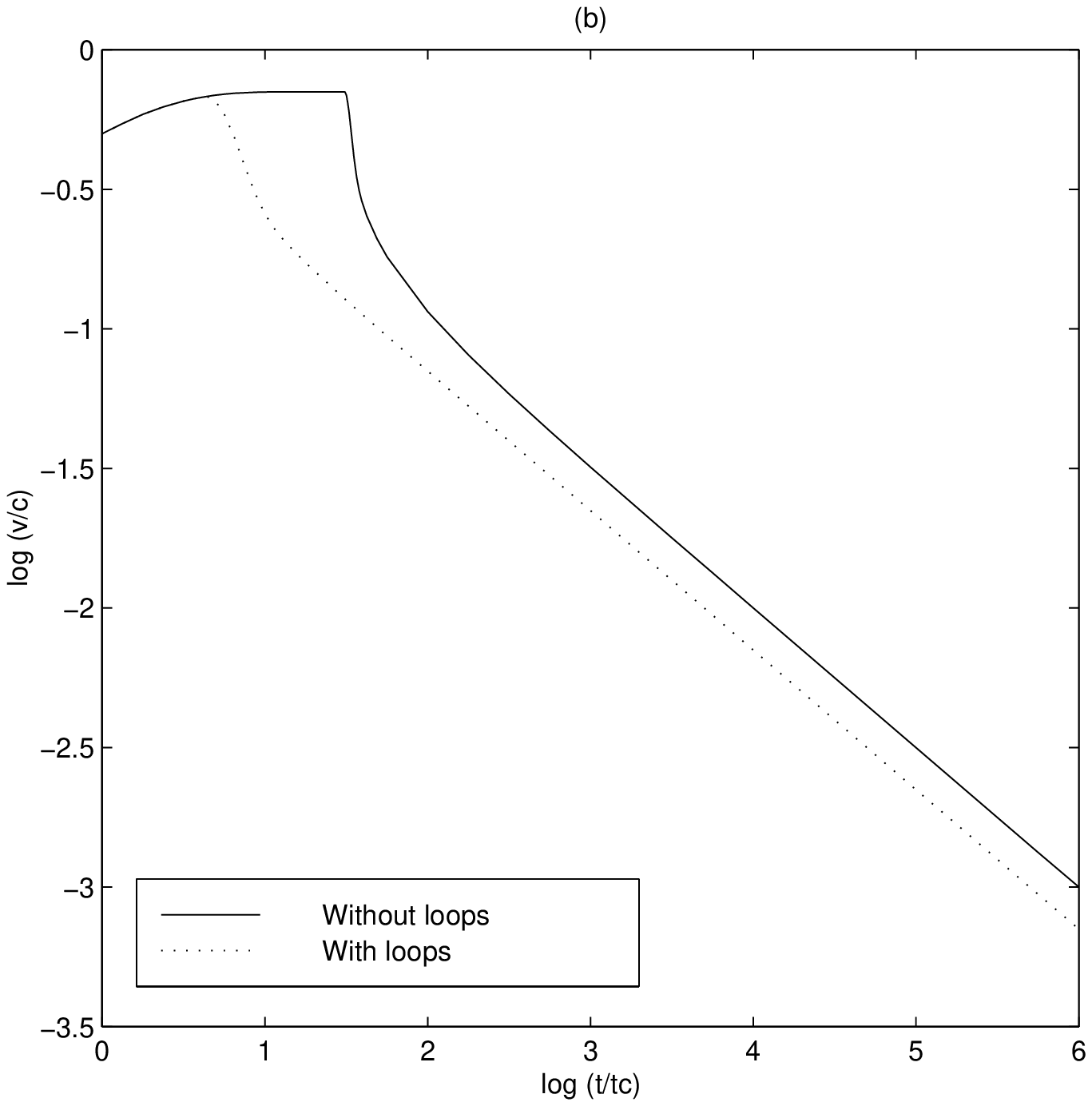}}
\vskip0.2in}
\vbox{\centerline{
\epsfxsize=.5\hsize\epsfbox{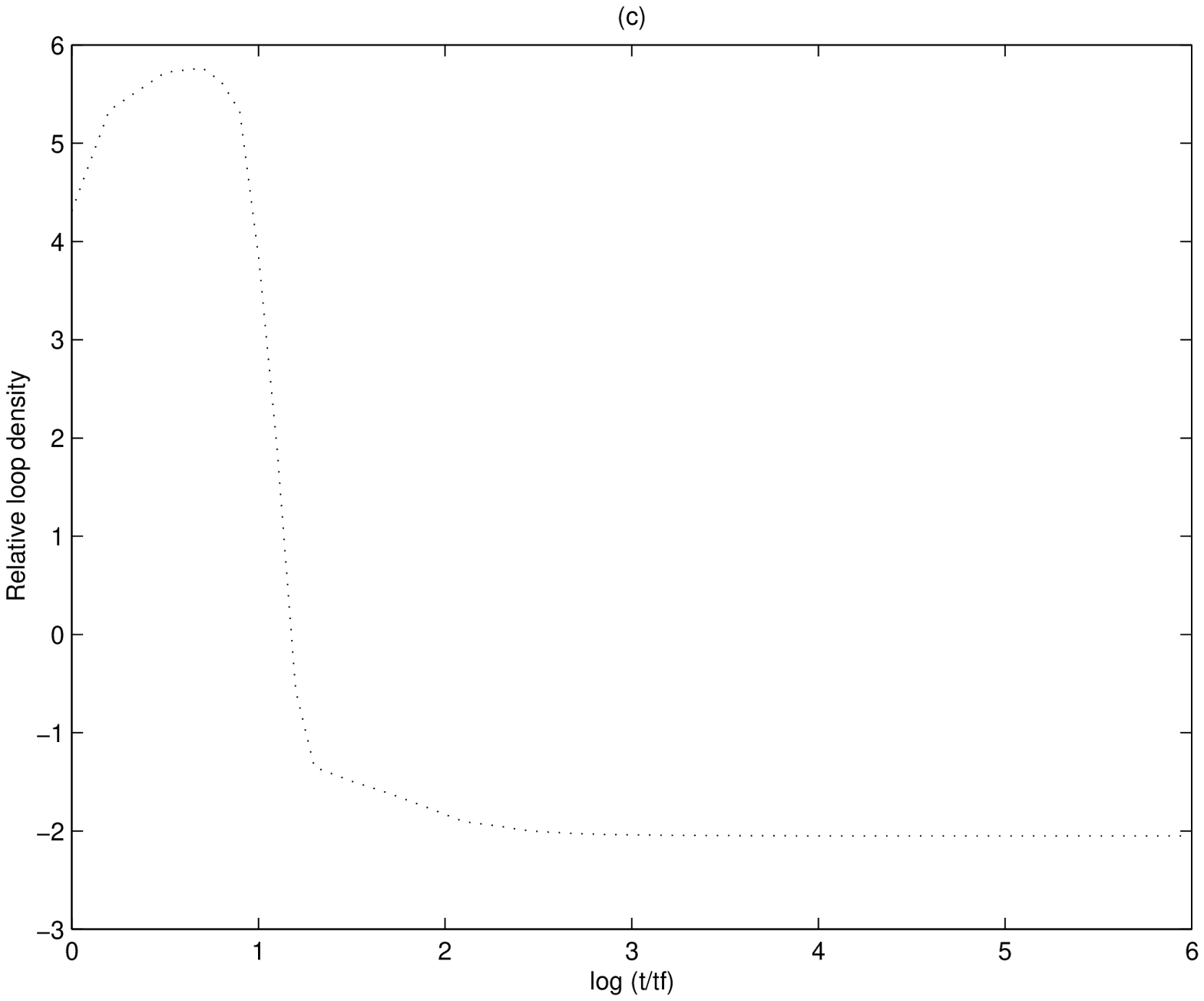}}
\vskip0in}
\caption{The evolution of an initially free network at
constant temperature, for ${\tilde c}=0$ (solid lines)
and ${\tilde c}=1$ (dotted lines). Plots represent the
lengthscale relative to the friction length (a),
velocity (in units of $c$)
(b) and ratio of long-string and loop energy densities
(c); all are log-log plots (time is in
orders of magnitude after the network formation).
Notice that the exponential growth in
$L$ (aided by the relativistic velocity) leads to a
period of loop domination.}
\label{fig46}
\end{figure}

Hence $L$ grows exponentially fast and quickly `catches
up' with $\ell_{\rm f}$; in other words,
a network starting in
the `free' regime rapidly evolves to the usual
friction-dominated regime. Figure \ref{fig46} displays
the evolution of a gauge string network in the free
regime, and in particular the transition to the damped
regime. Note that if loop production is allowed, this
fast growth of the correlation length will obviously
mean that an extremely large number of loops is produced.
In this case, the energy density in loops actually
exceeds the energy in long strings---in the above plot,
by a factor of $10^4$. A word of caution is however
needed here. In this case, loop reconnections onto the
long string network should play an important role.
However, since we still get an exponential growth if
loop production is switched off
(${\tilde c}=0$)---although the growth rate of $L$ is
obviously much larger for ${\tilde c}\not=0$---our
results should at least be qualitatively correct.

\section{Conclusions}
In this paper we have discussed the applicability
in condensed matter contexts of a
recently-developed model of string
evolution\cite{ms,ms1}, where a `characteristic
lengthscale' and the average velocity are the dynamical
variables. This has allowed us
to properly describe friction-dominated string dynamics,
hence providing the first complete and fully
quantitative study of string networks and their
corresponding loop populations in condensed matter
(as well as cosmological\cite{ms1}) contexts.
The fact that these results can be
obtained in a model initially aimed at describing
cosmic string evolution is, of course, a
manifestation of the universality of symmetry breaking
and defect formation phenomena, but it also lends
weight to the validity of this approach because
these cosmological models have been extensively 
tested numerically. 

We have confirmed two previously known
condensed matter scaling laws,
while noting the significant effects of loop production 
(which is
not considered in standard approaches). We also presented
the first (analytic or otherwise) discussion of the
evolution of the overall loop density, while also
determining loop lifetimes.
As we have already stated, the
literature on string loops in phase ordering
kinetics appears to be restricted to
a few references to the observation of loop
production and decay in laboratory experiments\cite{cdty}. It
is hoped that these more quantitative predictions
will trigger further efforts from the condensed
matter community.

\acknowledgments
We are grateful for the hospitality of the Isaac Newton
Institute where some of these problems were raised
during the {\it Topological Defects} workshop,
July--December, 1994. C.M.\ is funded by JNICT
(Portugal) under `Programa PRAXIS XXI' (grant no.
PRAXIS XXI/BD/3321/94). E.P.S.\ is funded by PPARC and
we both acknowledge the support of PPARC and the
EPSRC, in particular the Cambridge Relativity rolling
grant (GR/H71550) and a Computational Science
Initiative grant (GR/H67652).

\vfill\eject

\vfill\eject

\vfill\eject

\vfill\eject

\vfill\eject

\vfill\eject

\end{document}